\documentclass[aps,prd,superscriptaddress,amsmath,amssymb,onecolumn,nofootinbib]{revtex4-1}
\usepackage{amsthm,graphicx,multirow}
\usepackage{epsfig}
\usepackage{latexsym, amssymb} 
\usepackage{amsmath}
\usepackage{epstopdf}
\usepackage[perpage]{footmisc}
\usepackage{xcolor}

\usepackage{hyperref}

\def\beq{\begin{equation}}
\def\eeq{\end{equation}}
\def\br{\begin{eqnarray}}
\def\er{\end{eqnarray}}
\def\benu{\begin{enumerate}}
\def\efnu{\end{enumerate}}

\def\d{{\rm d}}

\begin{document}

\renewcommand{\arraystretch}{1.4}

\title{Cosmological parameter forecasts by a joint 2D tomographic approach to CMB and galaxy clustering}

\author{Jos\'e Ram\'on Bermejo-Climent}\email{jose.bermejo@inaf.it}
\affiliation{INAF OAS Bologna, via Piero Gobetti 101, Area della Ricerca CNR/INAF, I-40129 Bologna, Italy}
\affiliation{INFN, Sezione di Bologna, via Irnerio 46, I-40126 Bologna, Italy}
\affiliation{Departamento de Astrof\'isica, Universidad de La Laguna, 38206 La Laguna, Tenerife, Spain}

\author{Mario Ballardini}
\affiliation{Dipartimento di Fisica e Astronomia, Alma Mater Studiorum Universit\`a di Bologna, via Gobetti 93/2, I-40129 Bologna, Italy} 
\affiliation{INAF OAS Bologna, via Piero Gobetti 101, Area della Ricerca CNR/INAF, I-40129 Bologna, Italy}
\affiliation{INFN, Sezione di Bologna, via Irnerio 46, I-40126 Bologna, Italy}
\affiliation{Department of Physics \& Astronomy, University of the Western Cape, Cape Town 7535, South Africa}

\author{Fabio Finelli}
\affiliation{INAF OAS Bologna, via Piero Gobetti 101, Area della Ricerca CNR/INAF, I-40129 Bologna, Italy}
\affiliation{INFN, Sezione di Bologna, via Irnerio 46, I-40126 Bologna, Italy}

\author{Daniela Paoletti}
\affiliation{INAF OAS Bologna, via Piero Gobetti 101, Area della Ricerca CNR/INAF, I-40129 Bologna, Italy}
\affiliation{INFN, Sezione di Bologna, via Irnerio 46, I-40126 Bologna, Italy}

\author{Roy Maartens}
\affiliation{Department of Physics \& Astronomy, University of the Western Cape, Cape Town 7535, South Africa}
\affiliation{Institute of Cosmology and Gravitation, University of Portsmouth, Portsmouth PO1 3FX, United Kingdom}

\author{Jos\'e Alberto Rubi\~no-Mart\'in} 
\affiliation{Instituto de Astrof\'isica de Canarias, C/Via Lactea, s/n, 38205 La Laguna, Tenerife, Spain}
\affiliation{Departamento de Astrof\'isica, Universidad de La Laguna, 38206 La Laguna, Tenerife, Spain}

\author{Luca Valenziano}
\affiliation{INAF OAS Bologna, via Piero Gobetti 101, Area della Ricerca CNR/INAF, I-40129 Bologna, Italy}

\begin{abstract}
    The cross-correlation between the cosmic microwave background (CMB) fields and matter tracers carries important cosmological information. In this paper, we forecast by a signal-to-noise ratio analysis the information contained in the cross-correlation of the CMB anisotropy fields with source counts for future cosmological observations and its impact on cosmological parameters uncertainties, using a joint tomographic analysis. We include temperature, polarization and lensing for the CMB fields and galaxy number counts for the matter tracers. We consider $Planck$-like, the Simons Observatory, LiteBIRD and CMB-S4 specifications for CMB, and Euclid-like, Vera C. Rubin Observatory, SPHEREx, EMU and SKA1 for future galaxy surveys. We restrict ourselves to quasi-linear scales in order to deliver results which are free as much as possible from the uncertainties in modelling non-linearities. 
We forecast by a Fisher matrix formalism the relative importance of the cross-correlation of source counts with the CMB in the constraints on the parameters for several cosmological models. We obtain that the CMB-number counts cross-correlation can improve the dark energy Figure of Merit (FoM) at most up to a factor $\sim 2$  for LiteBIRD+CMB-S4 $\times$ SKA1 compared to the uncorrelated combination of both probes and will enable the Euclid-like photometric survey to reach the highest FoM among those considered here. We also forecast how CMB-galaxy clustering cross-correlation could increase the FoM of the neutrino sector, also  
enabling a statistically significant ($\gtrsim$ 3$\sigma$ for LiteBIRD+CMB-S4 $\times$ SPHEREx) detection of the minimal neutrino mass allowed in a normal hierarchy by using quasi-linear scales only.
Analogously, we find that the uncertainty in the local primordial non-Gaussianity could be as low as $\sigma (f_{\rm NL}) \sim 1.5-2$ by using two-point statistics only with the combination of CMB and radio surveys such as EMU and SKA1. Further, we quantify how cross-correlation will help characterizing the galaxy bias. 
Our results highlight the additional constraining power of the cross-correlation between CMB and galaxy clustering from future surveys which is mainly based on quasi-linear scales and therefore sufficiently robust to non-linear effects.
\end{abstract}

\maketitle

\section{Introduction}

Future large scale structure (LSS) surveys such as DESI\footnote{\href{https://www.desi.lbl.gov/}{https://www.desi.lbl.gov/}}, Euclid\footnote{\href{http://sci.esa.int/euclid/}{http://sci.esa.int/euclid/}}, SPHEREx\footnote{\href{http://spherex.caltech.edu/}{http://spherex.caltech.edu/}},  Vera C. Rubin Observatory, previously Large Synoptic Survey Telescope (LSST) \footnote{\href{http://www.lsst.org/}{http://www.lsst.org/}}, SKA\footnote{\href{http://www.ska.org/}{http://www.ska.org/}}, will provide maps of the matter tracers at different redshifts with an unprecedented sensitivity and sky coverage. These maps encode invaluable information on dark energy and modified gravity, dark matter and initial conditions 
which is complementary to the snapshot of the Universe taken by measuring the anisotropy pattern of the cosmic microwave background (CMB): the latter has been imaged with unprecedented sensitivity over all the sky by the Planck Collaboration with the third data release (PR3) in July 2018\footnote{\href{https://www.cosmos.esa.int/web/planck/planck}{https://www.cosmos.esa.int/web/planck/planck}}. The image of our Universe provided by the anisotropy pattern of the CMB has been generated at the cosmological recombination epoch occurred approximately 13.8 billion of years ago and during its propagation to the current epoch gets blurred by several intervening phenomena, such as gravitational lensing from structure formation, or the intervening change in time of the gravitational potentials, called integrated Sachs-Wolfe (ISW) effect, the Sunyaev-Zeldovich (SZ) effect and reionization. These effects (which are included among the so-called CMB secondary anisotropies) are correlated with the matter tracers at low redshift, such as galaxy/cluster counts or galaxy shear, whose surveys are expected to improve tremendously in the upcoming years.

The cross-correlation between CMB fields and galaxy surveys carries important cosmological information. A cross-correlation between the reconstructed lensing field from CMB temperature map and matter tracers has allowed the first detection of CMB lensing \cite{0705.3980,Hirata:2008cb}. The cross-correlation between temperature anisotropies and matter tracers has been used to measure at a statistical significant level the late-time integrated Sachs-Wolfe (ISW) effect \cite{astro-ph/9510072}, a contribution to CMB anisotropies on large angular scales, connected to the onset of the recent acceleration of the Universe \cite{Sachs:1967er,Kofman:1985fp}, which would be otherwise elusive in two point statistics of CMB temperature fluctuations. A detection of the ISW effect with an increasing statistical significance by the temperature-galaxy number density cross-correlation has been found through the years \cite{Nolta:2003uy,Ho:2008bz,Schiavon:2012fc,1303.5079,1502.01595,Stolzner:2017ged}.
Future galaxy survey promise an ISW detection with an higher significance \cite{Douspis:2008xv,Ballardini:2017xnt,1812.01636}. 

The cross-correlation between galaxy surveys and CMB lensing has an higher signal-to-noise compared to the corresponding one with CMB temperature \cite{Ade:2013tyw,Giannantonio:2015ahz,Omori:2018cid,Darwish:2020fwf}. Current estimates for this SNR is around $\sim 20$ (for $Planck$ and NVSS \cite{Ade:2013tyw}), 
but a sensible improvement is expected in the perspective of future CMB experiments which will improve the measurement of CMB lensing. It has been studied how this cross-correlation will contribute to the measurements of the amplitude of matter fluctuations, neutrino mass \cite{1311.0905,1710.09465}, primordial non-Gaussianities \cite{1710.09465,1802.08694,1906.04730} and galaxy bias. The cross-correlations of CMB lensing with galaxies and galaxy shear is also used in lensing ratio estimators, which can mitigate the uncertainties of the galaxy bias \cite{Das:2008am,Miyatake:2016gdc,Prat:2018yru,Bermejo-Climent:2019spz}.

More in general, the cross-correlation between CMB fields -temperature, polarization, lensing - and galaxy surveys will break degeneracies in cosmological parameters and will be the ideal completion of a fully combined CMB-LSS two dimensional (2D) tomographic likelihood \cite{Nicola:2016eua,Nicola:2016qrc,Doux:2017tsv}
using data of the next cosmological observations - just as the temperature-polarization $TE$ channel completes the joint CMB temperature-polarization 
likelihood.

In this paper, we forecast the impact of including the cross-correlation between the CMB fields 
and galaxies for a combined tomographic 2D analysis for current and future surveys with different specifications and characteristics. As CMB 
surveys we 
consider $Planck$\footnote{\href{https://www.cosmos.esa.int/web/planck}{https://www.cosmos.esa.int/web/planck}}, the Simons Observatory\footnote{\href{https://simonsobservatory.org/}{https://simonsobservatory.org/}} (SO), LiteBIRD\footnote{\href{http://litebird.jp/eng/}{http://litebird.jp/eng/}}, CMB-S4\footnote{\href{https://cmb-s4.org}{https://cmb-s4.org}}. As galaxy surveys we consider Euclid-like and Vera C. Rubin Observatory as examples for photometric surveys, Euclid-like spectroscopic as example for a pure spectroscopic survey, SPHEREx as example for a spectro-photometric survey, and EMU\footnote{\href{https://www.atnf.csiro.au/people/Ray.Norris/emu/index.html}{https://www.atnf.csiro.au/people/Ray.Norris/emu/index.html}} and SKA in Phase 1 (SKA1) as examples for radio continuum surveys. Note that in this paper we consider only the 2D tomographic clustering from Euclid-like photometric and spectroscopic surveys separately, and we refer to \cite{Ilicetal} for the combination and cross-correlation of the full Euclid capabilities \cite{Blanchard:2019oqi} with the CMB fields.

This analysis is performed for the current concordance $\Lambda$CDM model and some of its important extensions such as: 
(i) the $w_0,w_a$ parametrization for a dark energy component with a parameter of state dependent on the redshift, 
(ii) a neutrino sector in which $N_{\rm eff}$ and $\Sigma m_\nu$ are allowed to vary, (iii) primordial perturbations which allow the running of the spectral index and a local non-Gaussianity parameter which leads to a scale dependence for the galaxy bias. These generalizations beyond the $\Lambda$CDM cosmology in the dark energy, neutrino and primordial perturbation sectors are considered either separately, i.e. three different extensions of the concordance cosmology with 2 extra parameters, and jointly, i.e. as an extended cosmological model with 12 parameters (see \cite{DiValentino:2016hlg} for a study of a different 12 cosmological parameters model with current data), as an example of the type of extended cosmologies which could be studied in the future thanks to the improvement in the data and the combination between different kinds of data.

In terms of analysis, our study focus on the relevance of including CMB cross-correlation for multipoles where the linear perturbation is sufficiently adequate as in \cite{1207.6487}. This conservative 
cut to linear scales insure scientific validity to our analysis since accurate descriptions on non-linear scales are not available for all the cosmological models we analyze. At the same time it is useful to know how to use the whole cosmological information contained in linear scales, given the necessity to introduce theoretical uncertainties in correctly handling non-linear scales \cite{Baldauf:2016sjb}.

This paper is organized as follows: in Section \ref{sec:spectra} we define the quantities involved in our analysis, which are the 2D angular power spectra for the CMB, galaxy clustering and their cross-correlation. In Section \ref{sec:models} we discuss the cosmological models that we study. In Section \ref{sec:data} we describe the specifications adopted to model the mock data for the different CMB and galaxy surveys.
In Section \ref{sec:snr} we calculate the cross-correlation coefficients and signal-to-noise ratios of the temperature-galaxy and lensing-galaxy cross-correlations for the possible combinations of CMB and galaxy surveys. In Section \ref{sec:cosmoforecast} we describe the Fisher methodology used to forecast the parameter constraints. In Section \ref{sec:results} we discuss the constraints on the parameters of the various cosmological models, and in Section \ref{sec:conclusions} we draft our conclusions. In appendix \ref{sec:relativistic} we discuss the impact of the relativistic corrections on the galaxy counts and cross-correlation angular power spectra, and in appendix \ref{sec:constraints} we list the constraints on the cosmological parameters of all the models and combinations studied.

\section{CMB, galaxy counts and their cross-correlation}
\label{sec:spectra}

In our combined 2D tomographic analysis we consider the CMB temperature, polarization and lensing angular power spectra ($C_\ell^{TT}$, $C_\ell^{EE}$, $C_\ell^{\phi \phi}$), and their cross-correlations ($C_\ell^{TE}$, $C_\ell^{T \phi}$, $C_\ell^{E \phi}$), the galaxy counts auto spectra ($C_\ell^{GG}$) and its cross-correlation with the CMB fields. In this section, we focus on defining the cross-correlation of galaxy counts with the CMB temperature anisotropies ($C_\ell^{TG}$), and with the CMB lensing ($C_\ell^{\phi G}$), which are the most relevant, being $C_\ell^{E G}$ very small \cite{Challinor:2011bk}.

We compute the $C_{\ell}$ angular power spectra of the CMB temperature, polarization and lensing, of the galaxy clustering power spectra and all the cross-correlations between the different fields by a modified version of the publicly available code \texttt{CAMB\_sources}\footnote{\href{https://github.com/cmbant/CAMB/tree/CAMB\_sources}{https://github.com/cmbant/CAMB/tree/CAMB\_sources}} \cite{Challinor:2011bk}. Among our modifications we mention the specifications of each galaxy survey and the possibility of setting a scale-dependent bias.
 
The angular power spectra of the number counts and their cross-correlation with CMB are defined as
\begin{equation}
    C_{\ell}^{XY} = 4 \pi \int_0^{\infty} \frac{{\rm d}k}{k}\ {\cal P_R}(k) I_{\ell}^{X} (k) I_{\ell}^{Y} (k) \,,
\end{equation}
where $P_{\cal R}(k)$ is the dimensionless power spectrum of primordial curvature perturbation ${\cal R}$ and $I_\ell^X(k)$, $I_\ell^Y(k)$ are the kernels of the corresponding fields. 

The kernel of the galaxy 
number counts is given, if we neglect the corrections from redshift space distorsions (RSD) and general relativity (GR), by \cite{Challinor:2011bk}
\begin{equation}
\label{eq:kernel}
     I_{\ell}^{G} (k) = \int_0^{\infty} \frac{\d z}{(2\pi)^{3/2}} \frac{c}{H(z)} \frac{\d N}{\d z}(z) \Delta_k(z) j_\ell(k\chi(z)) \,,
\end{equation}
where $\d N/\d z$ is the normalized window function for the redshift distribution of sources, 
$\chi(z)$ is the conformal distance, and $j_\ell$ the spherical Bessel functions. Here
$\Delta_k(z)$ is the total number counts fluctuation in Newtonian gauge, usually approximated to 
$\Delta_k(z) \simeq b_G(z) \delta^{\rm c}_k(z)$ on sub-horizon scales, 
where $b_G(z)$ is the galaxy bias and $\delta^{\rm c}_k(z)$ is the comoving-gauge linear matter density perturbation. 
While lensing and other lightcone effects on the galaxy number counts angular power spectra 
have a small impact on the uncertainties on cosmological parameters, it will be necessary 
to model these contributions in order to avoid biases on cosmological parameters such as 
dark-energy parameters and the total neutrino mass \cite{Duncan:2013haa,Camera:2014bwa,Cardona:2016qxn,Lorenz:2017iez,1812.01636,Tanidis:2019fdh,Bermejo-Climent:2019spz}.
We summarize in appendix~\ref{sec:relativistic} all the RSD and GR contributions to $\Delta_k(z)$. 

The CMB temperature kernel, when cross-correlated with galaxy counts is given by the ISW contribution 
\begin{equation}
I_\ell^{\rm ISW}(k) = - \int_0^{\infty} \d z\ e^{-\tau}\left(\frac{\d \Phi}{\d z} + \frac{\d \Psi}{\d z}\right) j_\ell(k \chi(z))
\end{equation}
where $\tau$ is the reionization optical depth and $\Phi$, $\Psi$ are the gravitational potentials defined by the metric perturbations
\begin{equation}
\d s^2 = -(1+2\Psi) \d \eta^2 + (1-2\Phi) \d x^2 \,.
\end{equation}

The CMB lensing potential kernel is given by \cite{astro-ph/9611077} 
\begin{equation}
I_\ell^\phi(k)= \frac{3 \Omega_{m,0} H_0^2 }{  k^2 c} \int_0^{\infty} \frac{\d z}{(2\pi)^{3/2}} \frac{1+z}{H(z)}  \left(\frac{\chi(z_{*}) - \chi(z)}{\chi(z_{*}) \chi(z)}\right) \delta^{\rm c}_k(z) j_\ell(k\chi(z)) 
\end{equation}
 where $z_{*}$ is the redshift of the last scattering surface ($z_{*} \simeq 1100$). We discuss in Section \ref{sec:snr} the effect of the non-linear corrections on the CMB lensing, galaxy and cross-correlation power spectra, which are modelled with \texttt{halofit} \cite{1109.4416,1208.2701}.
 
\section{Cosmological models}
 \label{sec:models}
 
We adopt different 
 cosmologies as fiducial models in our analysis. 
 First, we test a $\Lambda$CDM cosmology and the $w_0$CDM model for the dark energy equation of state. We then consider three representative cases for the dark energy sector, neutrino physics and physics of the Early Universe, each modelled by a two parameters extension of the baseline $\Lambda$CDM cosmology. Finally, we consider as extCDM the 12 parameters cosmological model that considers jointly the three above mentioned extensions.

For the dark energy sector, we use the Chevallier-Polarski-Linder (CPL) parametrization \cite{gr-qc/0009008,astro-ph/0208512} of the parameter of state redshift dependence, given by 
\begin{equation}
\label{eq:de}
w(z) = w_0 + \frac{z}{1+z} w_a \,.
\end{equation}

For the neutrino physics we consider the minimal mass for the normal hierarchy assuming a total neutrino mass of $\Sigma m_\nu = 0.06$ eV in a single neutrino and two massless neutrinos. By allowing $N_{\rm eff}$ to vary, we consider the number of relativistic species (including massless neutrino) as the second free parameter.   

For the extensions connected to the physics of the Early Universe we consider a primordial power spectrum given by ${\cal P_R}(k) \equiv \frac{k^3}{2\pi^2} |{\cal R}_k|^2 =  A_s (k/k_*)^{n_s-1+ \frac{1}{2}\frac{\d n_s}{\d \ln k} \ln (k/k_*)}$ allowing $\frac{\d n_s}{\d \ln k} \neq 0$. 
Moreover, we consider a scale-dependent bias induced by a primordial local non-Gaussianity $f_{\rm NL}^{\rm loc}$ as \cite{Dalal:2007cu,Matarrese:2008nc}
  \begin{equation}
\label{eq:scalebias}
b(k,z) = b_G(z) + \Delta b(k,z) = b_G(z)+ [b_G(z)-1] f_{\rm NL}^{\rm loc} \delta_c \frac{3 \Omega_m H_0^2}{c^2 k^2 T(k) D(z)}
\end{equation}
where $b_G(z)$ is the usual linear galaxy bias calculated assuming  scale-independent Gaussian initial conditions, $\delta_c$ is the critical spherical overdensity ($\delta_c \simeq 1.686$ as predicted in \cite{Gunn:1972sv}), $T(k)$ is the matter transfer function, for which we adopt the analytical expression by \cite{astro-ph/9710252}, and $D(z)$ is the linear growth factor normalized according to the CMB convention. In Appendix~\ref{sec:neutrino} we discuss the impact of introducing also the scale dependent bias due to the neutrino mass, which is not considered in our baseline. Note that in this paper we constrain $f_{\rm NL}$ only by the scale-dependent effect in the galaxy bias function and we do not include the $f_{\rm NL}$ constraints from the CMB bispectrum \cite{1303.5079,1502.01595}. 

The fiducial cosmology is chosen consistently with $Planck$ 2018 results \cite{1807.06209}. We adopt $\Omega_b h^2$ = 0.022383, $\Omega_c h^2$ = 0.12011, $H_0$ = 67.32 km s$^{-1}$ Mpc$^{-1}$, $\tau$ = 0.0543, $n_s$ = 0.96605 and $\ln (10^{10} A_s)$ = 3.0448. For the non-standard parameters we assume as fiducial values $w_0$ = -1, $w_a$ = 0, $\Sigma m_\nu$ = 0.06 eV, $N_{\rm eff}$ = 3.046, $\d n_s
/ \d \ln k$ = 0 and $f_{\rm NL}$ = 0.
\section{Cosmological observations}
\label{sec:data}
In this section we describe the specifications of the CMB and galaxy surveys that we use for our signal and noise mock data.

\subsection{CMB surveys}

As CMB surveys, we use $Planck$-like synthetic data reproducing the $Planck$ 2018 results for $\Lambda$CDM parameters \cite{1807.06209}, the ground-based future experiments Simons Observatory (SO) \cite{1808.07445} and CMB Stage-4 (S4) \cite{1610.02743}, and Lite satellite for the studies of B-mode polarization and Inflation from cosmic background Radiation Detection (LiteBIRD) 
\cite{Hazumi:2019lys,Sugai:2020pjw}
as the next concept for a space mission dedicated to CMB polarization.

We consider the multipole coefficients $C_\ell^{TT} \,, C_\ell^{EE} \,, C_\ell^{TE}$ as signal for temperature, polarization and temperature-polarization cross-correlation, respectively.
As terms for the isotropic noise deconvolved with the instrument beam we consider \cite{astro-ph/9504054}
\begin{equation}
{\cal N}_{\ell}^X = w_X^{-1} b_{\ell}^{-2}, \qquad b_{\ell} = e^{-\ell(\ell+1)\theta_{\rm FWHM}^2/16 \ln 2}
\end{equation}
where $X = TT \,, EE$, $\theta_{\rm FWHM}$ is the full width half maximum (FWHM) of the beam in radians and $w_{TT}$, $w_{EE}$ are the inverse square of the detector noise level for temperature and polarization in arcmin$^{-1}$ $\mu$K$^{-1}$. For CMB lensing, we use the resulting ${\cal N}_{\ell}^{TT}$ and ${\cal N}_{\ell}^{EE}$ to reconstruct the minimum variance estimator for the noise ${\cal N}_{\ell}^{\phi\phi}$, combining the $TT$, $EE$, $BB$, $TE$, $TB$, $EB$ estimators according to the Hu-Okamoto algorithm \cite{astro-ph/0301031} and using the publicly available code \texttt{quicklens}\footnote{\href{https://github.com/dhanson/quicklens}{https://github.com/dhanson/quicklens}}. We show in Fig.~\ref{fig:cmb} the $TT$ and $\phi \phi$ power spectra and the corresponding noise for the experiments considered.

\subsubsection{$Planck$}
In order to reproduce a realistic simulation of $Planck$-like data we match our specifics in a manner that reproduce the $Planck$ 2018 results \cite{1807.06209}, which account for the entire data processing pipelines, including foreground contamination, systematics and other uncertainties that cannot be represented in our formalism. Therefore we consider only the 143 GHz channel with $w_{TT}$ = 33 $\mu$K arcmin, $w_{EE}$ = 70.2 $\mu$K arcmin and $\theta_{\rm FWHM} = 7.3$ arcmin and we inflate the noise in polarization, ${\cal N}_{\ell}^{EE}$, for $\ell < 30$ matching the resulting uncertainty of the optical depth in $Planck$ 2018 results. We consider the CMB lensing power spectrum ${C}_{\ell}^{\phi \phi}$ in the conservative range, i.e. for 
$8 \le \ell \le 400$, and neglect the $T \phi$, $E\phi$ cross-correlations according to the $Planck$ real likelihood.

\subsubsection{Simons Observatory}

The Simons Observatory \cite{1808.07445} will be a set of ground-based telescopes in Atacama, Chile, which is expected to have its first light in 2021. It will cover $\sim 40\%$ of the sky over six frequency bands ranging from 27 to 280  GHz, with a temperature sensitivity from 71 to 54 $\mu$K arcmin and a beam from 7.4 to 0.9 arcmin.
 To obtain ${\cal N}_{\ell}^{TT}$ and ${\cal N}_{\ell}^{EE}$ we combine the noise for the LAT baseline specifications of the six frequency bands given in \cite{1808.07445}. We re-adapt the resulting minimum variance lensing noise in order to match the baseline configuration in \cite{1808.07445}.

We assume $\ell_{\rm max}$ = 3000 for all the CMB channels. Since this is a ground-based experiment, we limit the minimum multipole to $\ell_{\rm min}$ = 40. We add the $Planck$-like specifications described above for $40 \leq \ell \leq 1500$ for the remaining sky fraction $f_\mathrm{sky}=0.3$ which is observed by $Planck$ but not by SO, and for $2 \leq \ell \leq 39$ with $f_\mathrm{sky}=0.7$ in order to include information from large scales.

\subsubsection{LiteBIRD}

 LiteBIRD \cite{Hazumi:2019lys,Sugai:2020pjw} is a proposal for a satellite selected by ISAS as a large strategic mission for the Japanese space agency (JAXA), with contributions of USA, Europe and Canada, with planned launch in 2027. Its main goal is the measurement of the CMB polarization anisotropy pattern with an angular resolution down to 18 arcmin and fifteen frequency channels spanning from 34 GHz to 448 GHz, a range optimized for the foreground removal. 
 As instrumental specifications for LiteBIRD, we use the 7 central frequency channels of the configuration described in \cite{Hazumi:2019lys}. We adopt $f_{\rm sky} = 0.7$ and $\ell_{\rm max}$ = 1350 as in \cite{Paoletti:2019pdi}. 

\subsubsection{CMB-S4}

CMB stage-4 \cite{1610.02743} describes the next generation CMB ground-based experiment. It will consist in a set of dedicated telescopes in the South Pole and Atacama. For S4 we adopt $w_{TT}$ = 1 $\mu$K arcmin, $w_{EE} = \sqrt{2}$ $\mu$K arcmin, $\theta_{\rm FWHM} = 3$ arcmin and $f_{\rm sky} = 0.4$ as in \cite{1509.07471}. Since S4 is ground-based, we use $\ell_{\rm min}$ = 40 and $\ell_{\rm max}$ = 3000, following \cite{1610.02743}. As for SO, we use complementary measurements in CMB temperature, polarization and lensing at large angular scales ($2 \le \ell \le 39$) and for the remaining fraction of the sky not observed by S4: given the timeline of CMB S4, for this purpose we use the capabilities of LiteBIRD.

\subsection{Galaxy surveys}
\label{sec:lss}
For future galaxy surveys we consider the ESA mission Euclid, the ground-based Vera C. Rubin Observatory  (LSST), the NASA mission Spectro-Photometer for the History of the Universe, Epoch of Reionization, and Ices Explorer (SPHEREx) and the
radio continuum galaxy surveys Evolutionary Map of the Universe (EMU) and Square Kilometer Array in Phase 1 (SKA1).

We parametrize the number density distribution of a galaxy survey $\d N/\d z$ as
\begin{equation}
\label{eq:dNdz}
\frac{\d N}{\d z} \propto f(z) \exp \left[ -\left( \frac{z}{z_0} \right)^{\beta} \right]
\end{equation}
where $f(z)$ is a redshift-dependent function and $z_0$, $\beta$ are parameters that depend on the galaxy survey.
We perform a tomographic analysis dividing the galaxy surveys in several redshift bins. For the Euclid-like, LSST and SPHEREx surveys we assume that the density distribution of a single bin is given by
\begin{equation}
\label{eq:binning}
\frac{\d n_{\rm gal}^i}{\d z} = \frac{\d N}{\d z} \int_{z_{\rm min}}^{z_{\rm max}}\d z_m p(z_m|z) 
\end{equation}
where $p(z_m|z)$ is the probability density for the measured redshift $z_m$ given the true redshift $z$ of the galaxy, and $z_{\rm min}$, $z_{\rm max}$ are the edges of the redshift bin, respectively. As baseline model for $p(z_m|z)$ we adopt a gaussian characterized by an intrinsic redshift scatter $\sigma_z$ as \cite{astro-ph/0506614}
\begin{equation}
\label{eq:photo-z}
p(z_m|z) = \frac{1}{\sqrt{2 \pi} \sigma_z} e^{-\frac{1}{2}(z_m - z)^2 / \sigma_z^2} \,.
\end{equation}
Solving Eq.~\eqref{eq:binning} we obtain
 \begin{equation}
\label{eq:densitybins}
\frac{\d n_i}{\d z} = \frac{1}{2} \frac{\d N}{\d z} \left[ {\rm erf} \left( \frac{z_{\rm max} - z}{\sqrt{2} \sigma_z } \right) - {\rm erf} \left( \frac{z_{\rm min} - z}{\sqrt{2} \sigma_z } \right) \right]
  \end{equation} 
where erf is the error function.

For the radio continuum galaxy surveys EMU and SKA1, since there is not a definition of the intrinsic scatter $\sigma_z$, we adopt gaussian windows with a dispersion equal to the half width of the bin according to the recipe in \cite{1810.06672}.  

\begin{figure}
\includegraphics[width = \columnwidth]{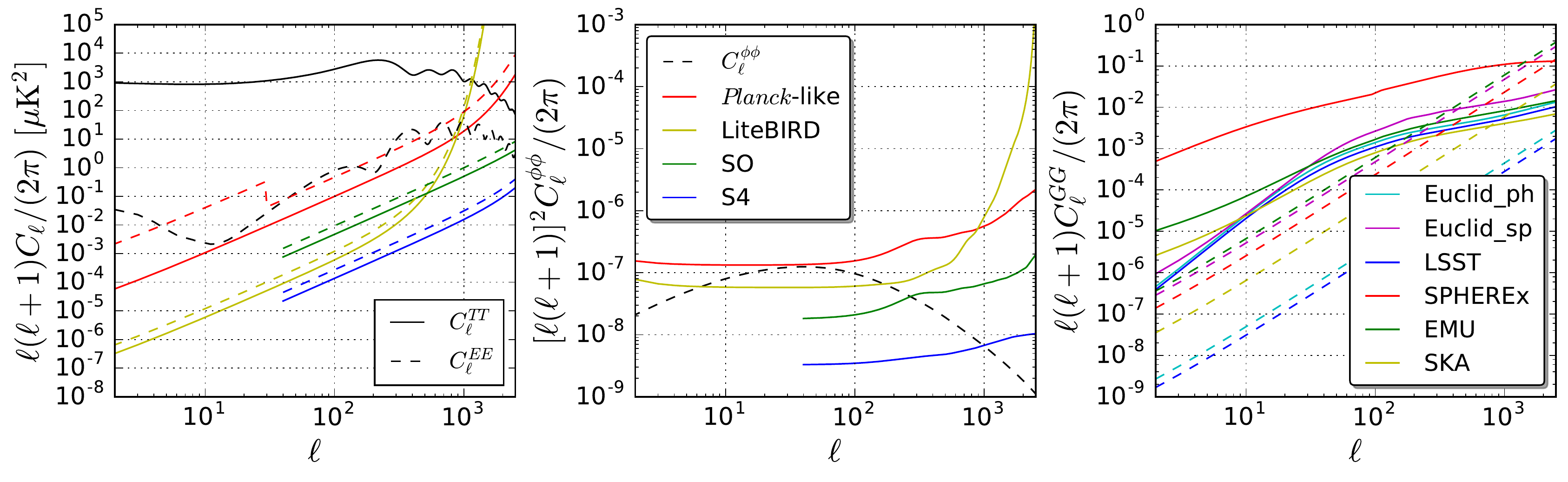}
\caption{Left panel: signal and noise for the CMB temperature and polarization for the various CMB surveys considered. The solid lines correspond to the temperature and the dashed lines to polarization. Mid panel: signal and noise for the CMB lensing. The black dashed curve corresponds to the $\phi \phi$ angular power spectra and the solid colored lines to the noise for the various CMB surveys. 
Right panel: signal and noise for the galaxy number counts. The solid lines represent the $GG$ angular power spectra of the single bin configuration for the 6 surveys, and the dashed lines the corresponding shot noise.}
\label{fig:cmb}
\end{figure}

The Poisson shot noise for the galaxy angular correlations between bins is given by
\begin{equation}
{\cal N}_{\ell}^{G}(z_i,z_j)  = \frac{\delta_{ij}}{n_{\rm gal}^{i}} 
\end{equation}
where $n_{\rm gal}^i$ is the number of objects per steradian unit in the $i$-th bin. We represent in Fig.~\ref{fig:cmb} the angular power spectra $C_\ell^{GG}$ for the single bin configuration of each survey and the corresponding shot noise.

\begin{figure}
\centering
  \includegraphics[width = 0.6\columnwidth]{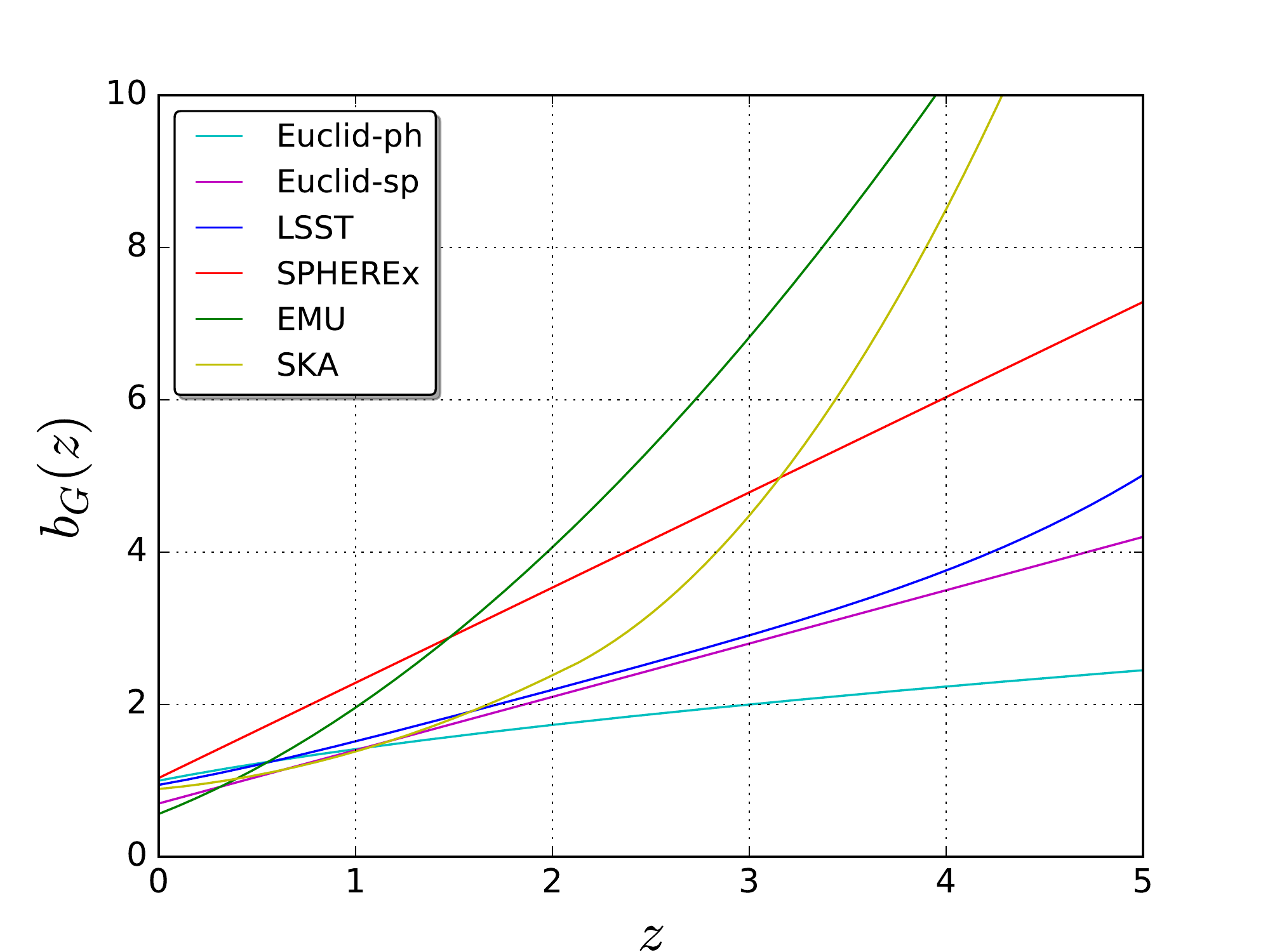}
\caption{Redshift evolution of the galaxy bias $b_G(z)$ for the six galaxy surveys considered.}
\label{fig:bz}
\end{figure}

For the galaxy bias redshift evolution, we adopt different functional forms for each survey, including a scale-dependent bias due to the primordial local non-Gaussianity contribution, as defined in Eq.~\eqref{eq:scalebias}. We represent in Fig.~\ref{fig:bz} the bias redshift evolution $b_G(z)$ for the four galaxy surveys.

\subsubsection{Euclid-like surveys}
The European Space Agency (ESA) Cosmic Vision mission Euclid \cite{1110.3193} is scheduled to be launched in 2022, with the goal of exploring the dark sector of the Universe. 
Euclid will measure the galaxy clustering in a spectroscopic survey of tens of millions of
H$\alpha$ emitting galaxies and the cosmic shear in a photometric survey of billions of galaxies.
We consider here the galaxy clustering from both surveys.

For the Euclid-like photometric survey (hereafter Euclid-ph-like) we parametrize the number density redshift distribution following Eq.~\eqref{eq:dNdz} with $f(z) = z^2$, $\beta$ = 3/2, $z_0 = 0.64$ and the distribution is normalized to account for number density of $\bar{n}_\mathrm{g}=30$ sources per arcmin$^2$. We consider a sky coverage of 15000 deg$^2$, and a redshift evolution of the bias following $b_G(z)=\sqrt{1+z}$ as in \cite{1206.1225}. For the tomographic analysis, we divide the survey in 10 redshift bins with the same number of sources per bin. The redshift accuracy is given by $\sigma_z = 0.05 (1+z)$. 

\begin{figure}
\includegraphics[width = \columnwidth]{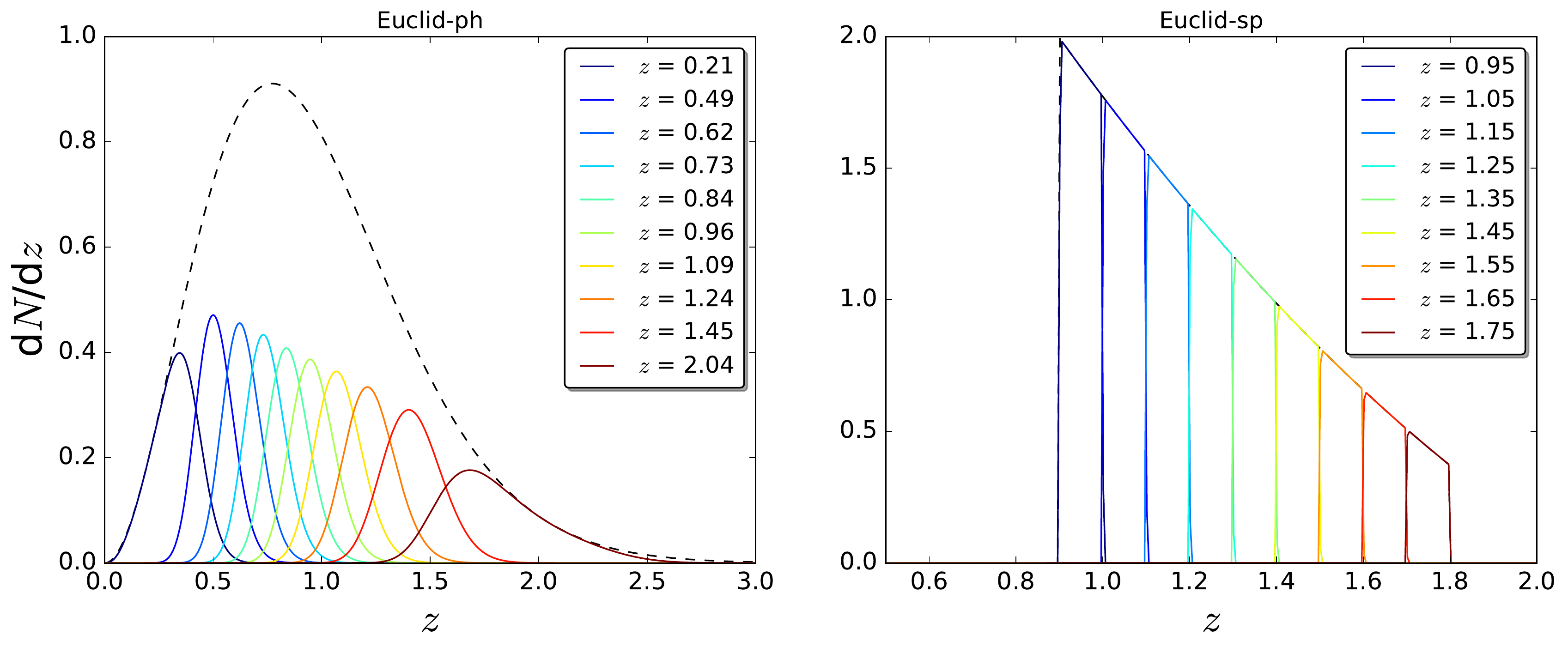}
\includegraphics[width = \columnwidth]{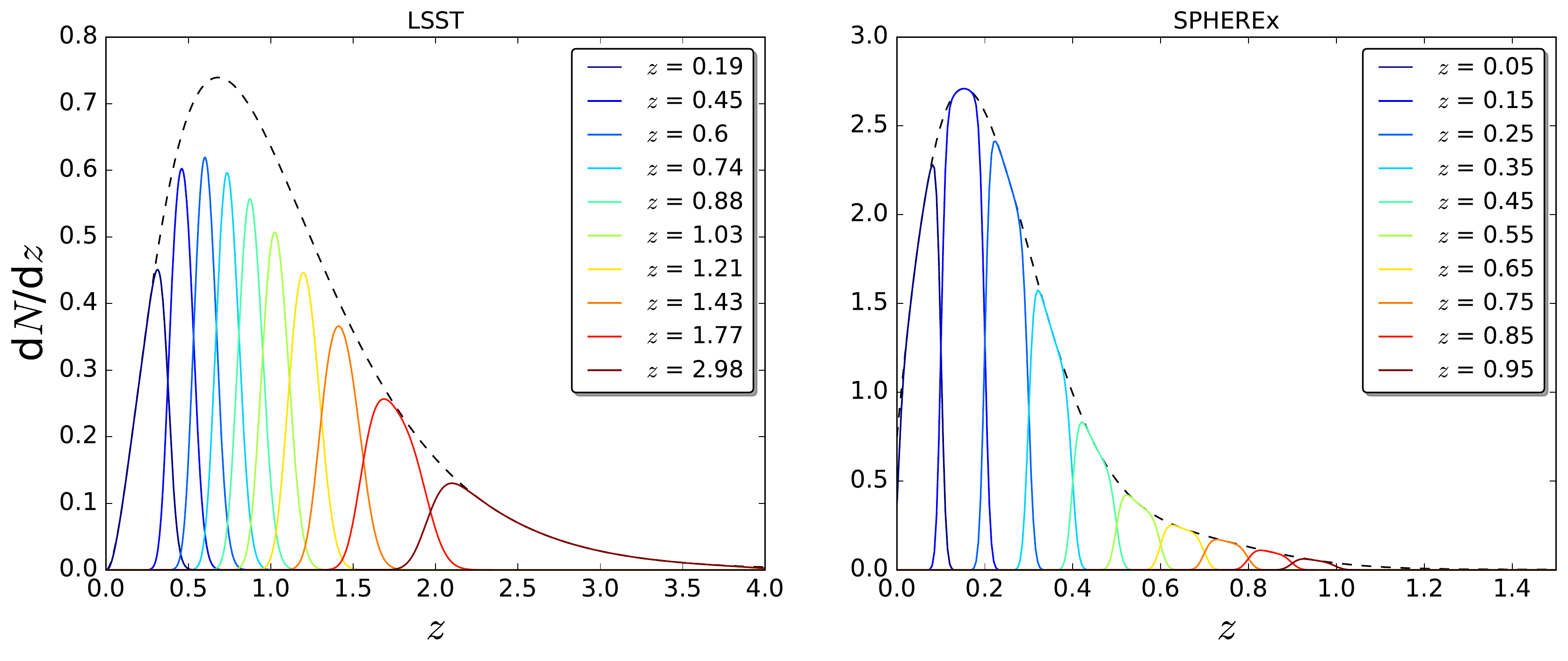}
\includegraphics[width = \columnwidth]{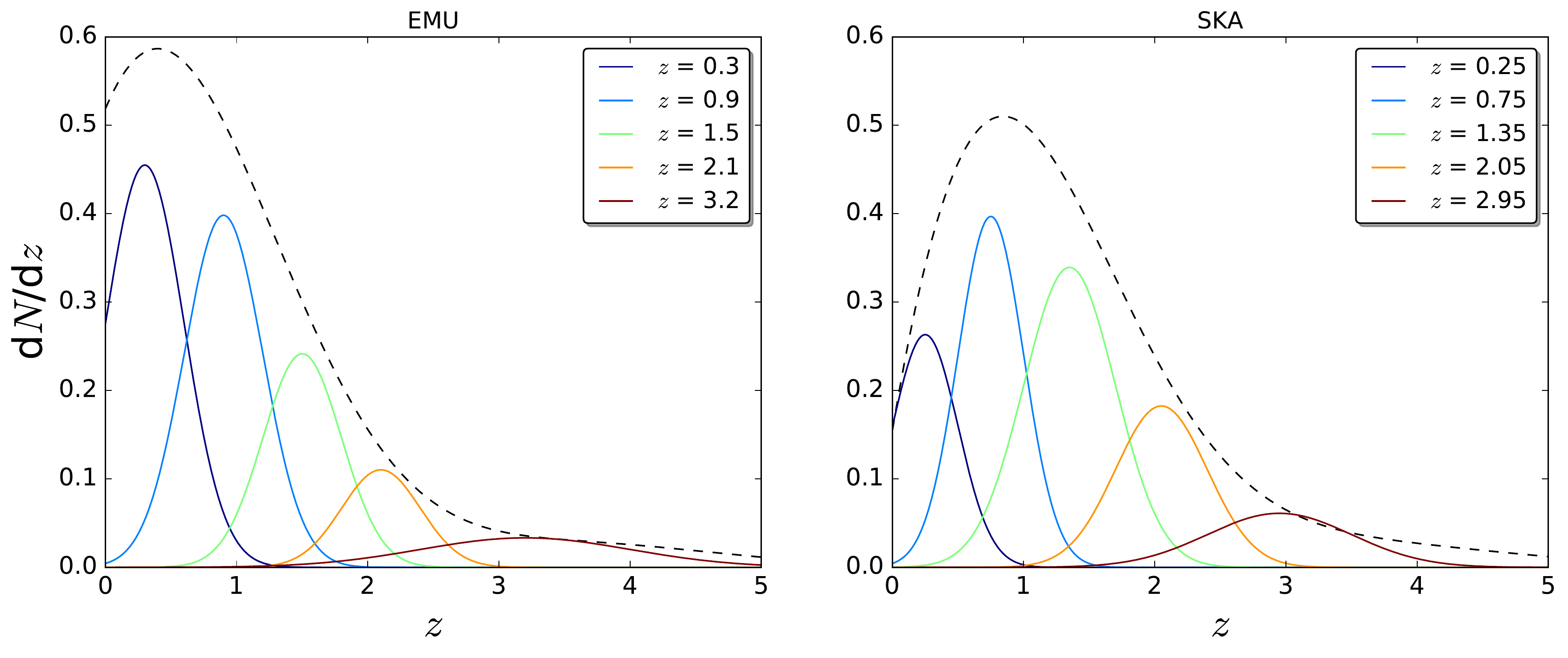}
\caption{Normalized number density of objects as a function of the redshift for our six surveys. The black dashed line represents the global density distribution $\d N/\d z$ of the survey and the colored lines represent the corresponding density distribution of the tomographic bins. We show also the central redshift of each bin.}
\label{fig:dNdz1}
\end{figure}

The Euclid-like spectroscopic survey (hereafter Euclid-sp-like) will measure the galaxy clustering from $\sim$ 30 million ${\rm H}\alpha$ emitters. According to the updated predictions obtained by 
\cite{Pozzetti:2016cch,1710.00833,Blanchard:2019oqi}, the Euclid-like wide single-grism survey will reach a flux limit 
$F_{\textup{H}\alpha} > 2 \times 10^{-16}\ \text{erg}\ \text{cm}^{-2}\ \text{s}^{-1}$ and will 
 cover a redshift range $0.9 \le z \le 1.8$. The sky coverage is also 15000 deg$^2$ and the expected number density of objects will be $\bar{n}_\mathrm{g} \simeq 2000$ sources per deg$^2$. We fit the number density distribution using the {\it model 3} data by \cite{Pozzetti:2016cch}, and assume as galaxy bias redshift evolution $b_G(z)= 0.7 + 0.7z$ following the fitting for emission line object from \cite{1903.02030}. For the tomography, we divide the survey in 9 bins with the same redshift width ($\Delta z = 0.1$), and the redshift accuracy is assumed to be $\sigma_z = 0.001 (1+z)$. In Fig.~\ref{fig:dNdz1} we show the normalized $\d N /\d z$ of both Euclid-like surveys and the binning choice.

\subsubsection{Vera C. Rubin Observatory}
The Vera C. Rubin Observatory, previously LSST, will be a 8.4 meter ground-based  telescope in Chile that will measure the galaxy clustering and weak lensing with a photometric survey that will cover a 13800 deg$^2$ area. This survey is expected to observe  $\bar{n}_\mathrm{g}=48$ sources per arcmin$^2$, with a linear galaxy bias which evolves with redshift following the equation $b_G(z)= 0.95/D(z)$ \cite{1809.01669}, where $D(z)$ is the linear growth factor. The number density redshift distribution is parametrized using Eq.~\eqref{eq:dNdz} with $f(z) = z^2$, $\beta$ = 0.9 and $z_{0}$ = 0.28. In the following we refer to Vera C. Rubin Observatory as LSST.

For the tomographic analysis, we divide the survey in 10 redshift bins with the same number of sources per bin. The redshift accuracy is given by $\sigma_z = 0.03 (1+z)$. In Fig.~\ref{fig:dNdz1} we show the normalized $\d N /\d z$ of the survey and the binning choice.

 \subsubsection{SPHEREx}
 SPHEREx is a recently approved NASA space mission that will measure the galaxy clustering using a spectro-photometric technique and covers $\sim 80\%$ of the sky \cite{1412.4872}. In this paper we assume the specifications of SPHEREx-2, a sample of the survey with low number density of galaxies but high redshift accuracy, this corresponds to $\sim 70$ million objects and $\sigma_z = 0.008(1+z)$. We fit the number density distribution following Eq.~\eqref{eq:dNdz} and the redshift evolution of the galaxy bias as in \cite{1606.03747}\footnote{We wish to thank Olivier Dor\'e and Roland de Putter for making available the SPHEREx specifications to us.}. For the tomography we divide the survey in 10 redshift bins with $\Delta z = 0.1$. We show in Fig.~\ref{fig:dNdz1} the number density distribution and the binning (note however that for SPHEREx the different flux cuts lead to different configurations).
 
\subsubsection{Radio continuum galaxy surveys}
\label{sec:radio}
Next generation of wide-area radio continuum surveys as SKA and one of its precursors, EMU, will play an 
important role in our understanding of the cosmology. In addition, wide-area radio continuum surveys will extend to substantially larger areas than many of 
the forthcoming optical surveys, i.e. $\gtrsim 20,000$ square degrees, and will not be affected by dust.
On the other hand, redshift information for 
individual sources can only be retrieved by the cross-correlation with surveys at other wavelengths \cite{Gomes:2019ejy}. 

EMU \cite{1106.3219} is an all-sky continuum survey planned for the new Australian SKA Pathfinder (ASKAP\footnote{\href{http://www.ska.org/}{https://www.atnf.csiro.au/projects/askap/}}). 
EMU will cover the same area (75\% of the sky) as NVSS, but will be 45 times 
more sensitive, and will have an angular  resolution (10 arcsec) five times better.
We consider for EMU a flux limit of 100 $\mu$Jy assuming a 10$\sigma$ r.m.s. detection 
threshold over the frequency range 1100-1400 MHz.

The Square Kilometer Array 1 (SKA1) is an international project to build a next generation telescope.
We study the SKA1-MID \cite{1811.02743} baseline: a dish array based in the Northern Cape province of South Africa.
We consider SKA1-MID in Band 1 covering slightly less than half the sky with a flux limit predicted to 
be at 25 $\mu$Jy assuming a 10$\sigma$ r.m.s. detection threshold over the frequency range 350-1050 MHz.

Radio continuum surveys will not provide redshift information, however it will be possible to separate sources by cross-correlating with other catalogues. Hence, we adopt a more conservative choice for the tomography and divide EMU and SKA1 in 5 broad bins in redshift as in \cite{1810.06672,1811.02743}. We represent the number density of sources and the binning in Fig.~\ref{fig:dNdz1}.

\section{Signal-to-noise analysis}

\label{sec:snr}
In this section we calculate the signal to noise ratio (SNR) of the cross-correlations of the CMB temperature and lensing\footnote{We do not show the SNR of the cross-correlation between CMB polarization and galaxies, $EG$, since it is small and noise dominated and very far from detection.} with galaxy number counts and study the impact of the tomography on this quantity. We assume the baseline $\Lambda$CDM cosmology defined in Section 3 and we also consider an alternative cosmology, with a larger a departure from $w=-1$ such as ($w_0, w_a$) = (-0.6, -1), 
in order to explore the dependence of the SNR on the fiducial model. We also check the importance of non-linear corrections.

\begin{figure}
\centering
\includegraphics[width = \columnwidth]{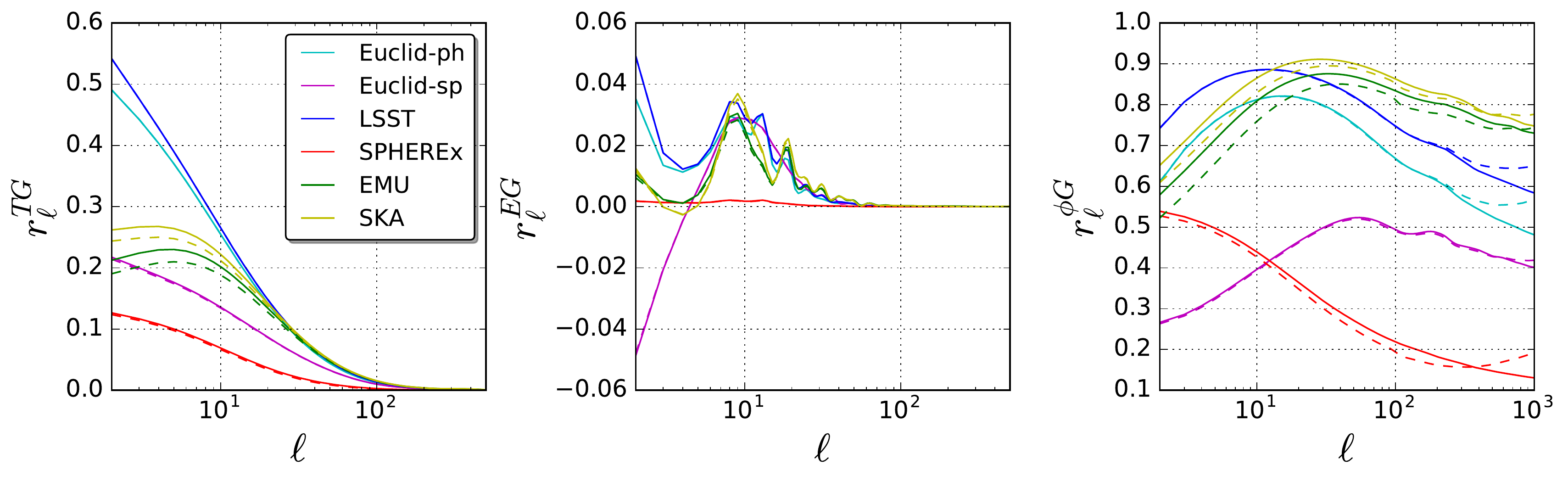}
\caption{Cross-correlation coefficients $r_{\ell}^{TG}$ (left panel), $r_{\ell}^{EG}$ (mid panel) and $r_{\ell}^{\phi G}$ (right panel) for the six galaxy surveys. The solid lines are obtained using the angular power spectra calculated with linear perturbations only, and the dashed lines include also the non-linear corrections from \texttt{halofit}.}
\label{fig:rl_nonlinear}
\end{figure}

We introduce the cross correlation coefficient as in \cite{1710.09465} 
\begin{equation}
r_{\ell}^{XG} \equiv \frac{C_{\ell}^{XG}}{\sqrt{C_{\ell}^{XX} C_{\ell}^{GG}}}
\end{equation}
where $X \equiv \{T,E,\phi\}$. 
In Fig.~\ref{fig:rl_nonlinear}, we show and compare the cross-correlation coefficients for the six galaxy surveys obtained using linear perturbations only and the same coefficients calculated including non-linear corrections from \texttt{halofit}, finding small differences between the two assumptions.

The SNR of the tomographic cross-correlation between a CMB field and galaxies is given by \cite{astro-ph/0611539}
\begin{equation}
\left( \frac{S}{N}\right)^2 = \sum_{i,j} \sum_{\ell = 2}^{\ell_{\rm max}} (2\ell+1)f_{\rm sky}^{XG} C_\ell^{XG}(z_i) [{\rm Cov}^{-1}_\ell]_{ij}C_\ell^{XG}(z_j)
 \label{SNR}
 \end{equation}
where the $i,j$ indices stand for the redshift bins, $X \equiv \{T,\phi\}$, $f_{\rm sky}$ is the overlapping sky fraction and Cov$_{\ell}$ is the covariance matrix, defined as
\begin{equation}
[{\rm Cov}_\ell]_{ij} = \bar{C}_\ell^{GG}(z_i,z_j) \bar{C}_\ell^{XX} + C_\ell^{XG}(z_i) C_\ell^{XG}(z_j)
\end{equation}
where $\bar{C}_\ell^{GG}(z_i,z_j) = C_\ell^{GG} (z_i,z_j) + \delta_{ij} {\cal N}_\ell^{GG}(z_i)$ and $\bar{C}_\ell^{XX} = C_\ell^{XX} + {\cal N}_\ell^{XX}$. For the cross-correlation sky fraction $f_{\rm sky}^{XG}$ we assume the common sky area to both CMB and galaxy surveys. We list in Tab.~\ref{tab:fsky} the overlapping sky fraction between each pair of CMB and galaxy surveys considered here.

\begin{table}
\centering
\caption{Overlapping sky fraction $f_{\rm sky}$ between each pair of CMB and galaxy surveys.}
\begin{tabular}{|l|cccccc|}
\hline
&  Euclid-ph-like & Euclid-sp-like & LSST & SPHEREx & EMU  & SKA1  \\ \hline

$Planck$ & 0.36 & 0.36 & 0.33 & 0.7 & 0.7 & 0.5 \\

SO  & 0.25 & 0.25 & 0.33 & 0.4 & 0.4 & 0.4  \\

LiteBIRD  & 0.36 & 0.36 & 0.33 & 0.7 & 0.7 & 0.5   \\

S4 &  0.25 & 0.25 & 0.33 & 0.4 & 0.4 & 0.4  \\
\hline 
\end{tabular} 
\label{tab:fsky}
\end{table}

As far as $\ell_{\rm max}$ is concerned in Eq.~\eqref{SNR}, we concentrate on quasi-linear scales in this paper. For $TG$ most of the information is concentrated only on linear scales and we cut the sum to $\ell_\mathrm{max}=200$.
For $\phi$G we adopt $\ell_{\rm max}^{\phi G} = \sqrt{\ell_{\rm max}^{\phi \phi} (\chi (\bar{z}) k_{\rm max} -1/2)}$, with $\ell_{\rm max}^{\phi \phi}$ = 1000, $\chi(\bar{z})$ is the comoving distance at the median redshift $\bar{z}$ of the redshift bin \cite{1207.6487,1809.07204} and $k_{\rm max}$ = 0.1 $h$/Mpc.

We calculate the SNR for  the $TG$ and $\phi G$ cross-correlations for all the combinations between the CMB and galaxy surveys considered. 
In order to quantify the impact of the tomography, we consider different configurations with a different number of bins starting from a single bin (i.e. the whole survey) up to the baseline number specified in Sect.~\ref{sec:lss}. We list in Tabs.~\ref{tab:TG} and \ref{tab:phiG} the SNR for $TG$ and $\phi G$, respectively, for the single bin case and the baseline number of bins. 

\begin{table}
\scriptsize
\centering
\caption{$TG$ signal-to-noise for the different combinations of CMB and galaxy surveys. The numbers between parenthesis correspond to the alternative fiducial model for ($w_0$, $w_a$).}
\begin{tabular}{|l|cccccccccccc|}
\hline
&  \multicolumn{2}{c}{Euclid-ph-like} & \multicolumn{2}{c}{Euclid-sp-like} & \multicolumn{2}{c}{LSST} & \multicolumn{2}{c}{SPHEREx} & \multicolumn{2}{c}{EMU}  & \multicolumn{2}{c|}{SKA1}  \\
 & 1 bin & 10 bins &  1 bin & 9 bins & 1 bin & 10 bins   & 1 bin & 10 bins & 1 bin & 5 bins & 1 bin & 5 bins \\ \hline 

$Planck$ &  3.8 (4.3) & 4.0 (4.4) & 2.2 (2.5) & 2.2 (2.5) & 3.8 (4.3) & 4.0 (4.4) &  1.3 (1.4) & 4.2 (4.8) & 4.3 (4.8) & 5.0 (5.6) & 4.1 (4.5) & 4.6 (5.1)  \\

$Planck$+SO &  3.8 (4.3) & 4.0 (4.4) & 2.2 (2.5) &  2.2 (2.5) & 3.8 (4.3) & 4.0 (4.4)  & 1.3 (1.4) &  4.4 (5.0) & 4.3 (4.8) & 5.0 (5.6) & 4.1 (4.6) & 4.7 (5.2) \\

LiteBIRD+S4 &  3.8 (4.3) & 4.0 (4.4) & 2.2 (2.5) &  2.2 (2.5) & 3.8 (4.3) & 4.0 (4.4) & 1.3 (1.4) &  4.4 (5.0) & 4.3 (4.8) & 5.0 (5.6) & 4.1 (4.6) & 4.7 (5.2) \\
\hline 
\end{tabular} 
\label{tab:TG}
\end{table}

\begin{table}
\scriptsize
\centering
\caption{$\phi G$ signal-to-noise for the different combinations of CMB and galaxy surveys. The numbers between parenthesis correspond to the alternative fiducial model for ($w_0$, $w_a$).}
\begin{tabular}{|l|cccccccccccc|} 
\hline
 &\multicolumn{2}{c}{Euclid-ph-like} & \multicolumn{2}{c}{Euclid-sp-like}  & \multicolumn{2}{c}{LSST} &\multicolumn{2}{c}{SPHEREx} & \multicolumn{2}{c}{EMU}  & \multicolumn{2}{c|}{SKA1} \\

& 1 bin & 10 bins & 1 bin & 9 bins & 1 bin & 10 bins & 1 bin & 10 bins & 1 bin & 5 bins & 1 bin & 5 bins \\ \hline

$Planck$  & 61 (59) & 73 (71) & 43 (42) & 43 (42) & 67 (65) & 82 (80) & 21 (20) & 54 (52) & 80 (79) & 88 (87) & 87 (85) & 93 (92) \\

$Planck$+SO  &  95 (92) & 119 (116) & 71 (70) & 71 (70) & 118 (115) & 152 (150) & 28 (27) & 89 (86) & 116 (115) & 132 (131) & 145 (143) & 161 (160) \\

LiteBIRD+S4  & 137 (133) & 181 (178) & 103 (102) & 108 (107) & 152 (149) & 208 (206) & 38 (37) & 138 (133) & 175 (174) & 216 (215) & 201 (200) & 239 (238) \\
\hline
\end{tabular}
\label{tab:phiG}
\end{table}

For either the baseline cosmology and the alternative one, we obtain similar SNRs. However, the SNR for $TG$ can increase around $\sim$ 10-15\% by assuming the alternative values of ($w_0$, $w_a$), demonstrating the capability of the CMB temperature-galaxy  cross-correlation to help discriminating among deviations from $\Lambda$. 
Since the CMB temperature anisotropy pattern is signal dominated for the multipoles relevant for the $TG$ cross-correlation already with $Planck$, the SNR is mostly dependent on the common sky fraction of  the CMB map and the galaxy survey. For the baseline fiducial cosmology we find that next photometric surveys from Euclid and LSST will reach a $4 \sigma$ detection for the ISW, but the radio continuum galaxy surveys are the most promising in this respect, reaching $\sim$ 5$\sigma$ in the EMU and SKA1 tomographic configurations, consistently with  \cite{1812.01636}. We get a slightly better ISW detection with EMU as a consequence of its sky fraction, that allows a larger overlap with the CMB, as shown in Tab.~\ref{tab:fsky}. We also note that SPHEREx is the survey which benefits more from a tomographic approach for the ISW detection.

\begin{figure}
\centering
 \includegraphics[width = \columnwidth]{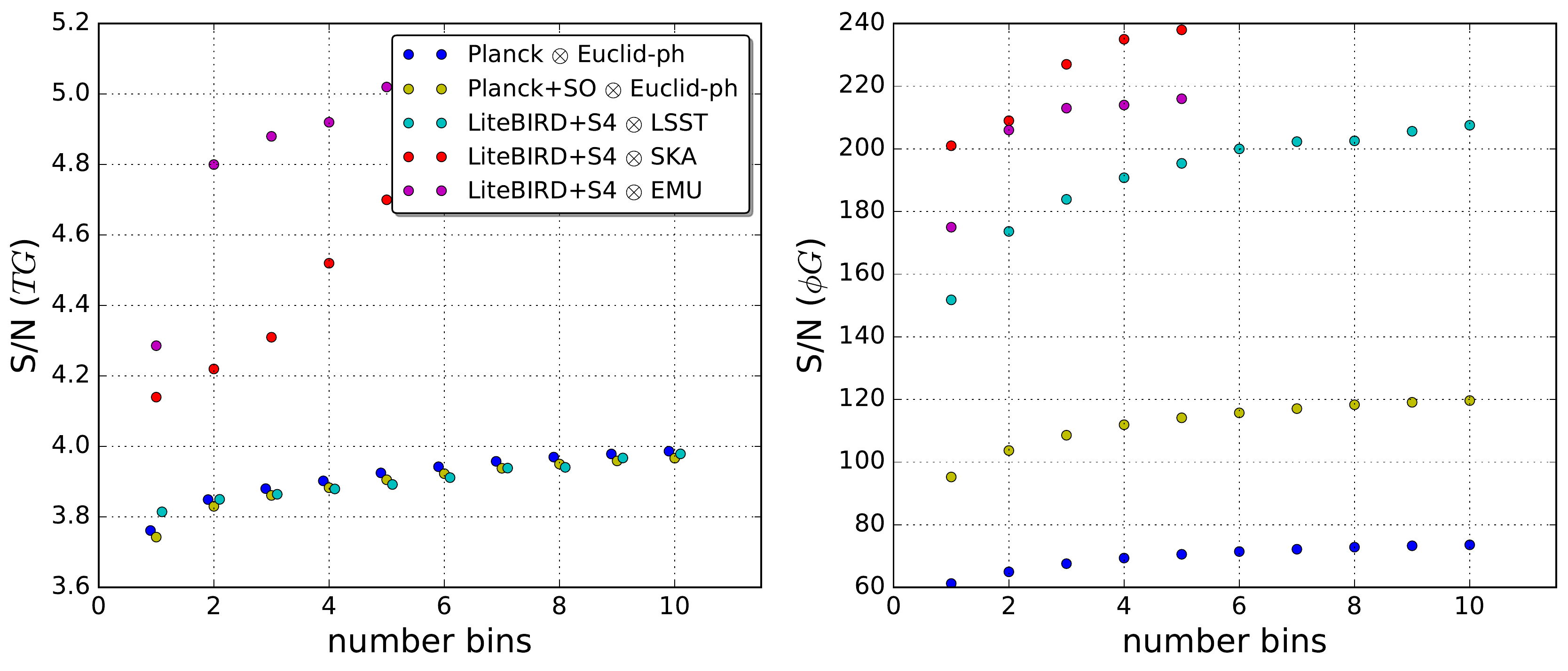}
\caption{Signal-to-noise ratio of the $TG$ (left panel) and $\phi G$ (right panel) cross-correlations as a function of the number of bins $N$. The blue dots correspond to $Planck$ $\otimes$ Euclid-ph-like, the yellow dots to $Planck$+SO $\otimes$ Euclid-ph-like, the cyan dots to LiteBIRD+S4 $\otimes$ LSST, the red dots to LiteBIRD+S4 $\otimes$ SKA1 and the purple dots to LiteBIRD+S4 $\otimes$ EMU.}
\label{fig:SNR_bins}
\end{figure}

In the case of $\phi G$, there is a larger margin of improvement expected with the next CMB polarization experiments with respect to the current $\sim 20 \sigma$ detection obtained by $Planck$ and NVSS \cite{Ade:2013tyw}.
For all the future galaxy surveys considered we indeed obtain a larger SNR.
For $\phi G$ the CMB lensing noise is dominant for $Planck$ whereas it decreases significantly for $Planck$+SO and LiteBIRD+S4 as shown in Fig.~\ref{fig:cmb}. As a consequence, the SNR for $\phi G$ can increase up to a factor $\sim$ 2-3 for LiteBIRD+S4 with respect to $Planck$. Once again, SPHEREx is the survey which benefits more from a tomographic approach also for the CMB lensing-galaxy cross-correlation. We also note that the values for the SNRs obtained for the alternative cosmology are very similar to the fiducial one, albeit slightly smaller.

We show in Fig.~\ref{fig:SNR_bins} the behavior of the $TG$ and $\phi G$ SNR as a function of the number of bins $N$ to complement the information contained in Tabs.~\ref{tab:TG} and \ref{tab:phiG}. We consider the combinations of $Planck$ and $Planck$+SO with Euclid-like photometric survey, and LiteBIRD+S4 with LSST, EMU and SKA1, and divide the galaxy surveys progressively in different number of bins up to the baseline (10 bins for Euclid-ph-like and LSST and 5 bins for EMU and SKA1). We obtain that the $TG$ SNR for Euclid-ph-like and  LSST saturates around $\sim 4\sigma$ independently of the CMB survey chosen, as can be already seen from Tabs.~\ref{tab:TG} and \ref{tab:phiG}. This is a consequence of the increasing Poisson shot noise and redshift overlapping between bins when pushing the tomography. For $\phi G$, we get an improvement of the SNR as a consequence of future better measurement of CMB lensing, and we find as well a saturation when increasing the number of bins for both Euclid-like and LSST: the better SNR reached by LSST compared to the Euclid-like photometric survey is due to the overlap in the scanning strategy of SO and LSST. We finally note that a more aggressive binning scheme than the one adopted here could enhance the scientific capability of EMU and SKA.

\section{Fisher analysis for cosmological parameter forecasts}
\label{sec:cosmoforecast}
We use the Fisher matrix information \cite{astro-ph/9603021} to forecast cosmological parameters uncertainties. In the Fisher formalism, the likelihood ${\cal L}$ is assumed to be a multivariate Gaussian and the minimum errors on the cosmological parameters are given by the diagonal of the inverse of the Fisher matrix as $\sigma_i \geq \sqrt{({\cal F}^{-1})_{ii}}$. The Fisher matrix ${\cal F}$ is defined as:
\begin{equation}
\label{eq:fishertrace}
{\cal F}_{\alpha \beta} = \left\langle \frac{\partial^2 {\cal L}}{\partial \theta_\alpha \partial \theta_\beta} \right\rangle = \frac{1}{2} {\rm Tr} \left[  \frac{\partial {\cal C}}{\partial \theta_\alpha} {\cal C}^{-1} \frac{\partial {\cal C}}{\partial \theta_\beta}  {\cal C}^{-1} \right]
\end{equation}

where ${\cal C}$ is the theoretical covariance matrix and $\theta_\alpha$, $\theta_\beta$ are the cosmological parameters. If we take into account the number of modes given by $(2\ell+1) f_{\rm sky}/2$, Eq.~\eqref{eq:fishertrace} becomes

\begin{equation}
\label{eq:fishertrace2}
{\cal F}_{\alpha \beta} = \sum_{\ell_{\rm min}}^{\ell_{\rm max}} \sum_{abcd} \frac{2 \ell+1}{2} f_{\rm sky}^{abcd} \frac{\partial {C_\ell^{ab}}}{\partial \theta_\alpha} ({\cal C}^{-1})^{bc} \frac{\partial {C_\ell^{cd}}}{\partial \theta_\alpha} ({\cal C}^{-1})^{da}
\end{equation}
where $abcd$ $\in$ $\{T,E,\phi,G_1, ..., G_N \}$ \footnote{Note that we consider $C_{\ell}^{TT} \,, C_{\ell}^{TE} \,, C_{\ell}^{EE}$ to avoid double counting the lensing contribution as in \cite{Errard:2015cxa}.} and $f_{\rm sky}^{abcd} \equiv \sqrt{ f_{\rm sky}^{ab}f_{\rm sky}^{cd} }$ is the effective sky fraction for each pair of channels. The theoretical covariance matrix ${\cal C}$ is defined as
\begin{equation}
\label{eq:cov}
{\cal C} = 
\begin{bmatrix}
   \bar{C}_{\ell}^{TT} & C_{\ell}^{TE} & C_{\ell}^{T\phi} & C_{\ell}^{TG_1} & \dots  & C_{\ell}^{TG_N} \\
    C_{\ell}^{TE} &    \bar{C}_{\ell}^{EE} & C_{\ell}^{E\phi} & C_{\ell}^{EG_1} & \dots  & C_{\ell}^{EG_N} \\
        C_{\ell}^{T\phi} & C_{\ell}^{E\phi} &    \bar{C}_{\ell}^{\phi\phi} & C_{\ell}^{\phi G_1} & \dots  & C_{\ell}^{\phi G_N} \\
     C_{\ell}^{TG_1} & C_{\ell}^{EG_1} & C_{\ell}^{\phi G_1} &    \bar{C}_{\ell}^{G_1 G_1} & \dots  & C_{\ell}^{G_1 G_N} \\
    \vdots & \vdots & \vdots & \vdots & \ddots & \vdots \\
     C_{\ell}^{TG_N} & C_{\ell}^{EG_N} & C_{\ell}^{\phi G_N} & C_{\ell}^{G_1 G_N} & \dots  &    \bar{C}_{\ell}^{G_N G_N} \end{bmatrix} \,.
\end{equation}

We include in our analysis nuisance parameters to account for the uncertainties in the number density distribution and galaxy clustering bias of the surveys. We vary the redshift parameter $z_0$ to consider the uncertainties on the ${\rm d}N/{\rm d}z$ function, and include a free constant per bin as $b_n(\bar{z}_n)b_G(z)$, where $n$ is the $n$-th redshift bin, 
to account for the uncertainties on the galaxy clustering bias function. In the single bin cases, we also include an extra nuisance parameter that modifies the slope of $b_G(z)$. Hence, we include 3 nuisance parameters for the single bin cases and $N$+1 nuisance parameters for the $N$ bins tomographic cases. 

Concerning the minimum multipole $\ell_{\rm min}$ in Eq.~\eqref{eq:fishertrace}, we use as baseline $\ell_{\rm min}^{GG}=\ell_{\rm min}^{TG}=\ell_{\rm min}^{\phi G}$. 
We link $\ell_{\rm min}^{GG}$ to the sky fraction covered by the galaxy survey and we adopt $\ell_{\rm min}^{GG}$ = 10 for Euclid-like, $\ell_{\rm min}^{GG}$ = 20 for LSST, $\ell_{\rm min}^{GG}$ = 5 for SKA. We consider $\ell_{\rm min}^{GG}$ = 2 for SPHEREx and EMU since both cover a wider sky fraction. For $\ell_{\rm max}^{\phi G}$ we restrict to quasi-linear scales as discussed in Section \ref{sec:snr} and we set 
$\ell_{\rm max}^{G G} = \chi (\bar{z}) k_{\rm max} -1/2$ with $\chi(\bar{z})$ is the comoving distance at the median redshift $\bar{z}$ of the redshift bin \cite{1207.6487,1809.07204} and $k_{\rm max}$ = 0.1 $h$/Mpc. 

\begin{figure}
\centering
\includegraphics[width = \columnwidth]{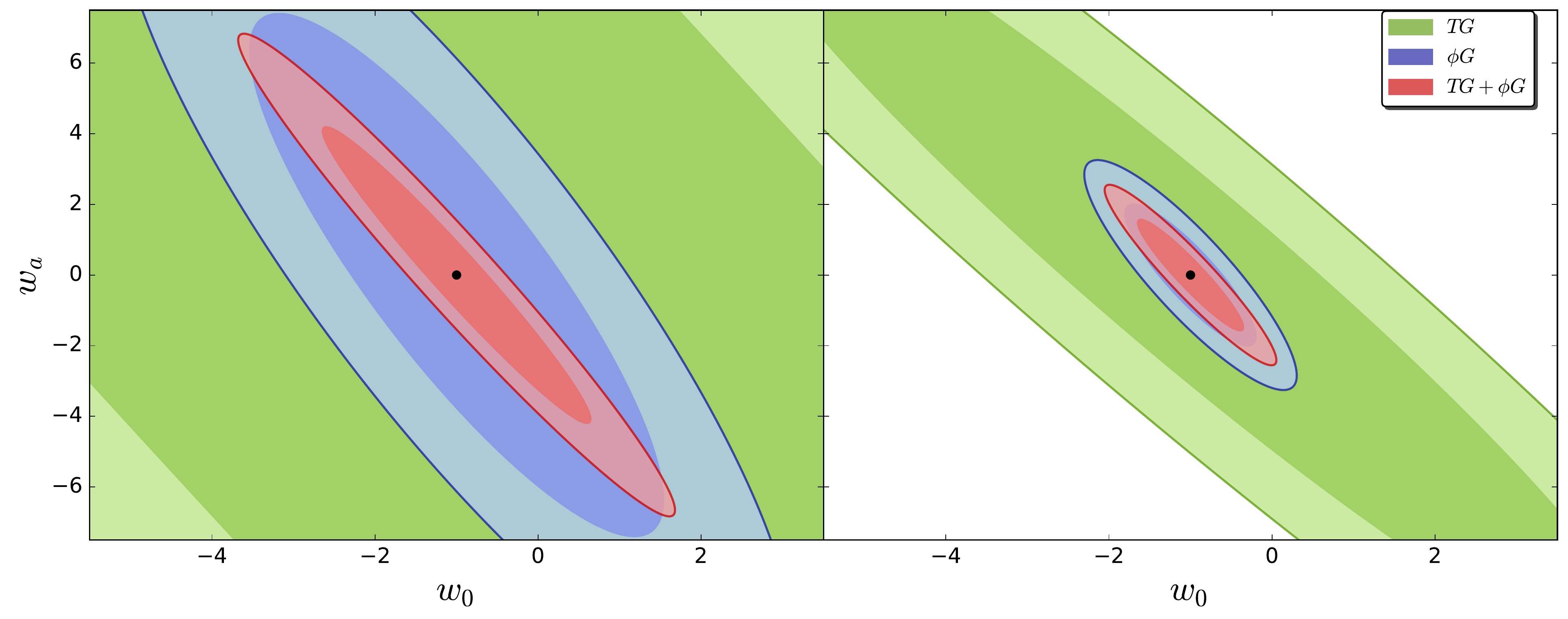}
\caption{Marginalized 68\% and 95\% 2D confidence regions for the constraints from $TG$, $\phi G$ and the combination of both for the $w_0 w_a$CDM model. The left panels correspond to $Planck$ $\otimes$ Euclid-ph-like and the right panels to LiteBIRD+S4 $\otimes$ SKA1. The green contours correspond to the temperature-galaxy cross-correlation constraints ($TG$) the blue contours to the lensing-galaxy cross-correlation ($\phi G$) and the red contours to the sum of both ($TG+\phi G$).}
\label{fig:ellipses2}
\end{figure}

In order to quantify the relevance of including the CMB-galaxy cross-correlation for the parameter constraints, we compare the forecast uncertainties using two approaches. The first one is the simple combination of CMB and galaxy clustering (GC) as uncorrelated probes, which we call CMB $\oplus$ GC. In this case, we apply Eq.~\eqref{eq:fishertrace2} to both probes independently and we add the two resulting Fisher matrices, being equivalent to neglecting the off-diagonal blocks in Eq.~\eqref{eq:cov} that account for the CMB-galaxy cross-correlation. In the second approach, which we call CMB $\otimes$ GC, we compute one joint Fisher matrix using the full covariance matrix described by Eq.~\eqref{eq:cov} that includes the $TG$, $EG$ and $\phi G$ correlations. We choose to present the results in this way since the CMB-GC cross-correlation alone would lead to loose constraints on parameters. Fig. \ref{fig:ellipses2} indeed shows
the constraints from $TG$ and $\phi$G separately and jointly for the dark energy extension for $Planck$ $\otimes$ LSST and LiteBIRD + CMB-S4 $\otimes$ SKA1. The information from $\phi G$ provides better constraints than $TG$ alone, but the combined constraints $TG$+$\phi G$ are still loose.

It is also useful to introduce the Figure of Merit (FoM) to quantify the capability of constraining a pair of parameters ($\alpha$,$\beta$) as \cite{1206.1225}
\begin{equation}
{\rm FoM}_{\alpha,\beta} =\frac{1}{ \sqrt{\det({\cal F}^{-1}_{\alpha,\beta} )}} \,, 
\label{eq:fom}
\end{equation}
where ${\cal F}^{-1}_{\alpha,\beta} $ is the 2x2 covariance matrix of the two parameters. More generally, we can define the FoM of $N$ parameters as \cite{1609.08510,1610.09290}
 \begin{equation}
{\rm FoM}_{\alpha_i} =\left[\frac{1}{ \det({\cal F}^{-1}_{\alpha_i} ) }\right]^{1/N}
\label{eq:fomgeneral}
\,,
  \end{equation}
where ${\cal F}^{-1}_{\alpha_i} $ is the covariance $N$x$N$ corresponding  matrix of the parameters. We use the first definition to calculate the FoM of two parameters; we use the second one for the FoM of the primary cosmological parameters and for the FoM of the bias nuisance parameters of a given model. Note that the FoM defined in \cite{DiValentino:2016foa} can be obtained by $(\cdot)^{N/2} $ the one in Eq.~\eqref{eq:fomgeneral}.

 \section{Results}
 \label{sec:results}
  \begin{figure}
\centering
\includegraphics[width = \columnwidth]{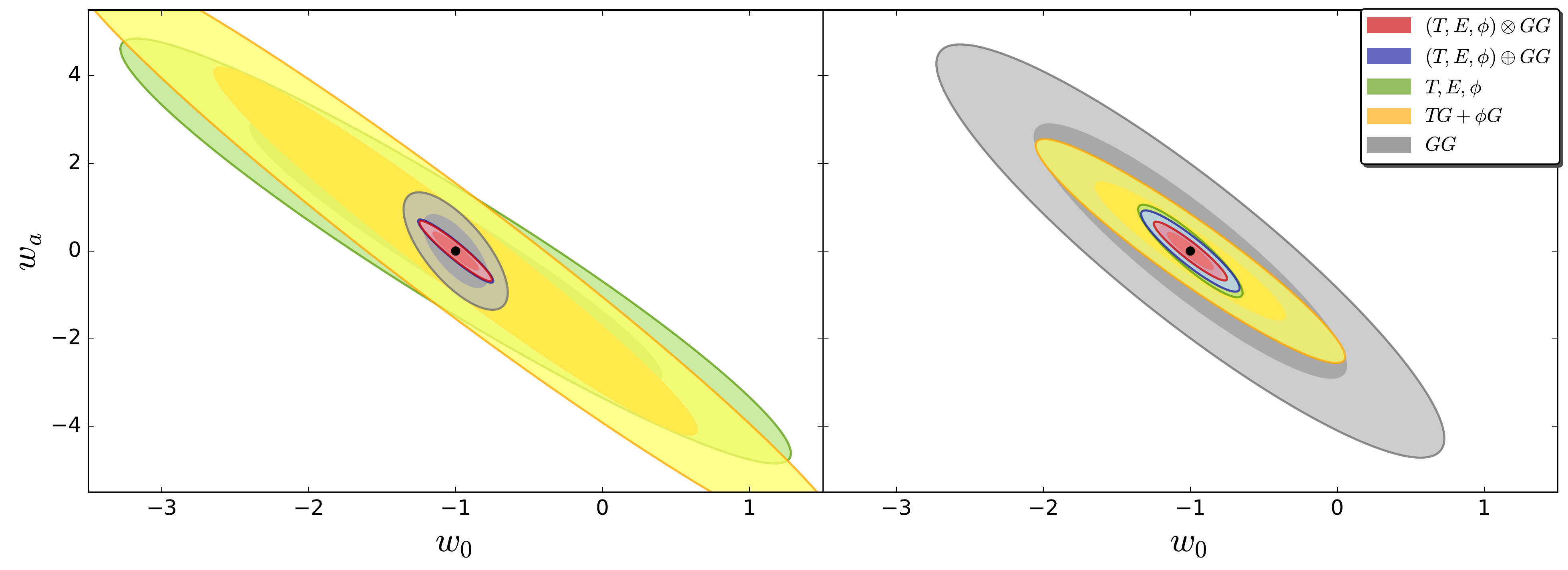}
 \includegraphics[width = \columnwidth]{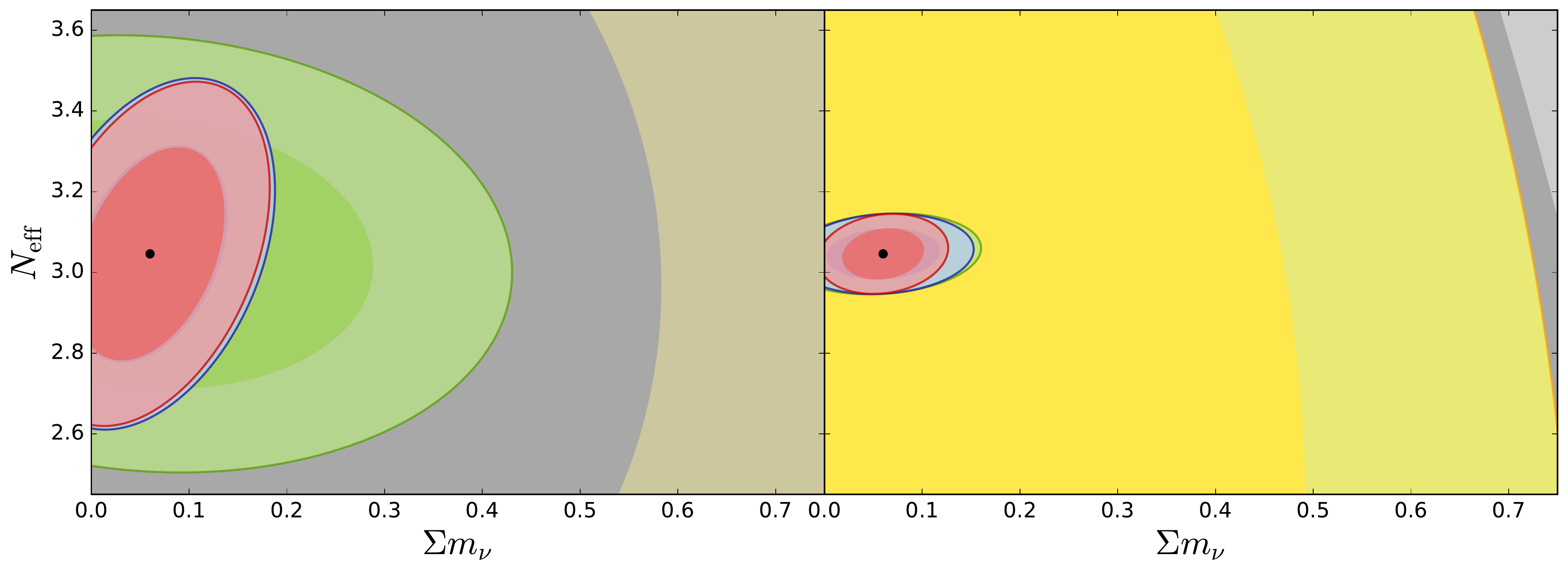}
\includegraphics[width = \columnwidth]{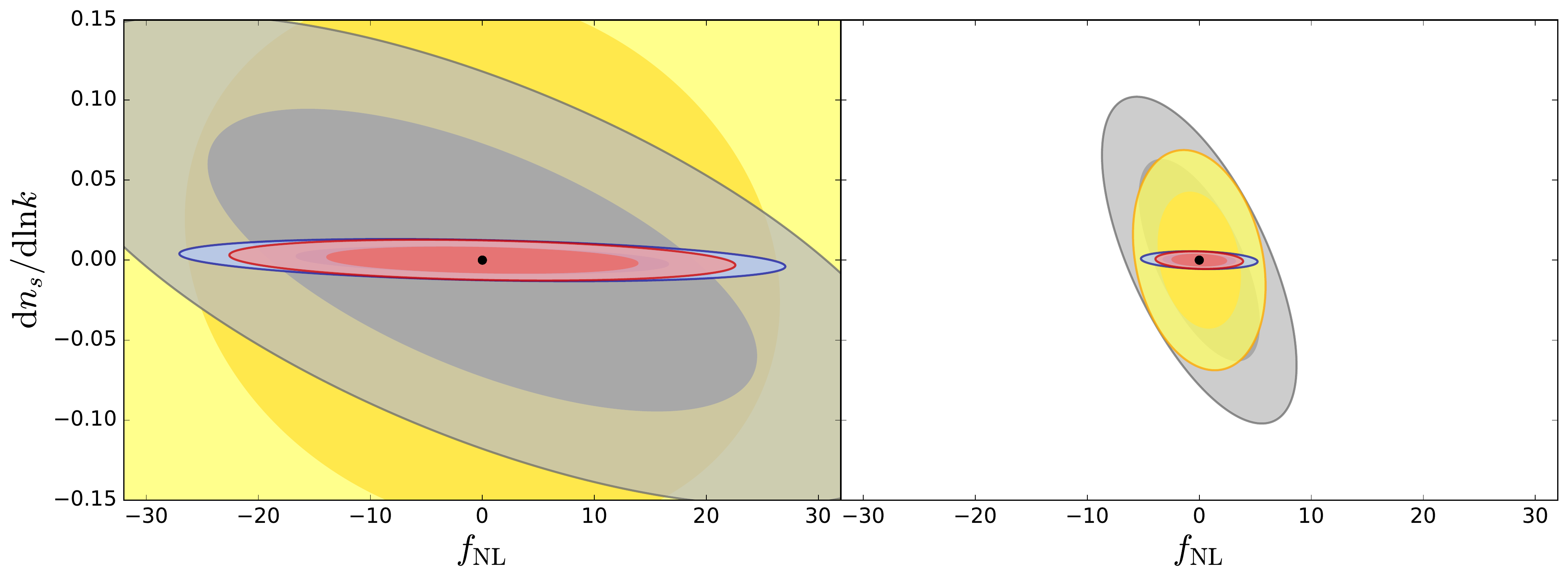}

\caption{Marginalized 68\% and 95\% 2D confidence regions for the constraints from CMB, GC and cross-correlation independently and jointly for the three 2 parameter $\Lambda$CDM extensions: dark energy (top), neutrino physics (middle) and primordial universe (bottom). The left panels correspond to $Planck$ $\otimes$ Euclid-ph-like and the right panels to LiteBIRD+S4 $\otimes$ SKA1. The green contours correspond to the CMB-only constraints ($T,E,\phi$), the yellow contours to the cross-correlation only ($TG,\phi G$), the grey contours to the galaxy counts ($GG$), the blue contours to the CMB-GC combination as uncorrelated ($T,E,\phi \oplus GG$) and the red contours to the combination including cross-correlation ($T,E,\phi \otimes GG$). In the bottom panel the green contours are not shown since $f_{\rm NL}$ is not constrained from the CMB information in our analysis.}
\label{fig:ellipses}
\end{figure}

 We quantify the relevance of the CMB-GC cross-correlation comparing the parameter constraints obtained with the CMB $\oplus$ GC and CMB $\otimes$ GC approaches described in Section \ref{sec:cosmoforecast}. 
In order to evaluate the impact of tomography, for each galaxy survey we present results either by using a single redshift bin and the baseline number of bins discussed in Section \ref{sec:lss}.
We first discuss the $\Lambda$CDM model, then $w_0$CDM, the three two parameters extensions - $w_0 w_a$CDM, $\Lambda$CDM + $\{\Sigma m_\nu, N_{\rm eff}\}$, $\Lambda$CDM + $\{\d n_s/ \d \ln k$, $f_{\rm NL}\}$ - and then the 12 parameters extCDM model. For each model in the Tabs. of Appendix B, we quote the 68\% marginalized uncertainties on cosmological parameters and FoMs, which show the improvement in cosmology and in the characterization of the galaxy clustering bias. 

\begin{figure}
\centering
 \includegraphics[width = \columnwidth]{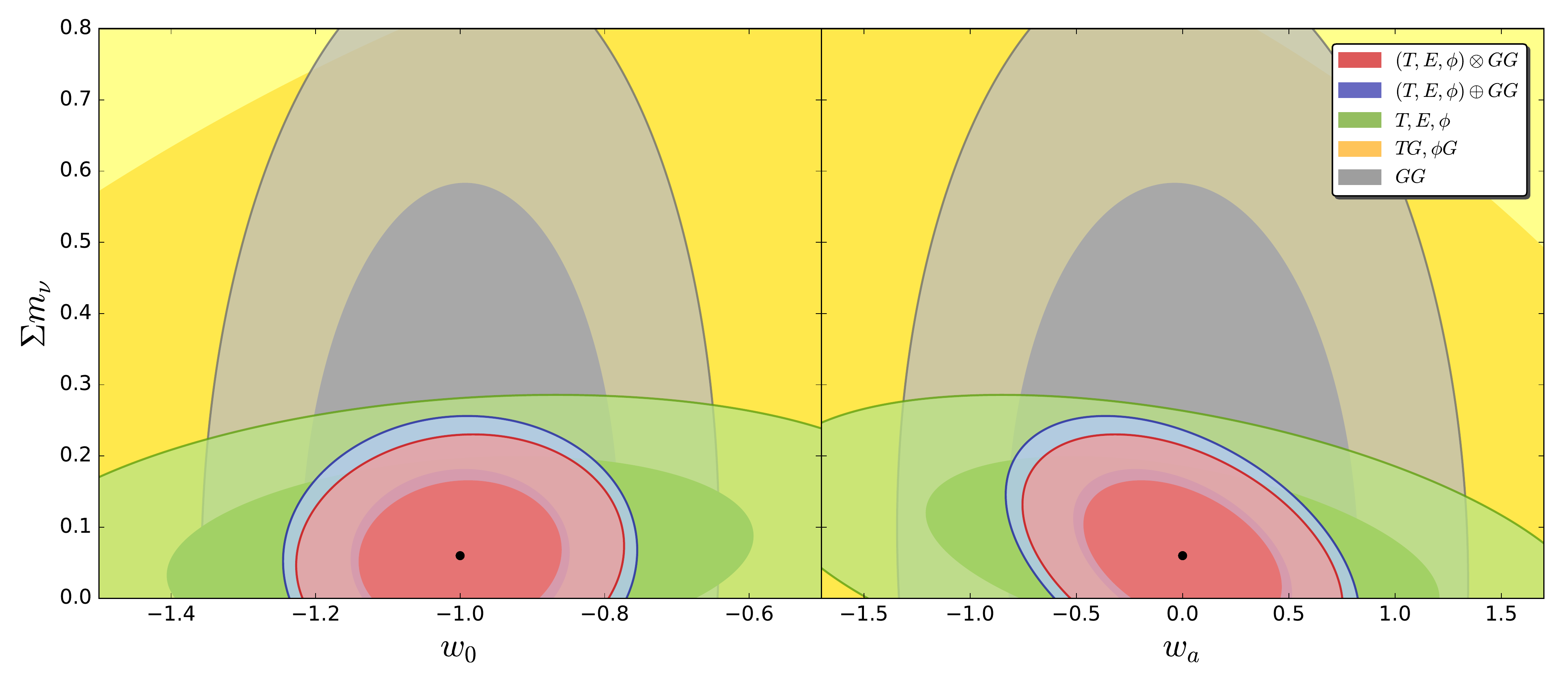}
 \includegraphics[width = \columnwidth]{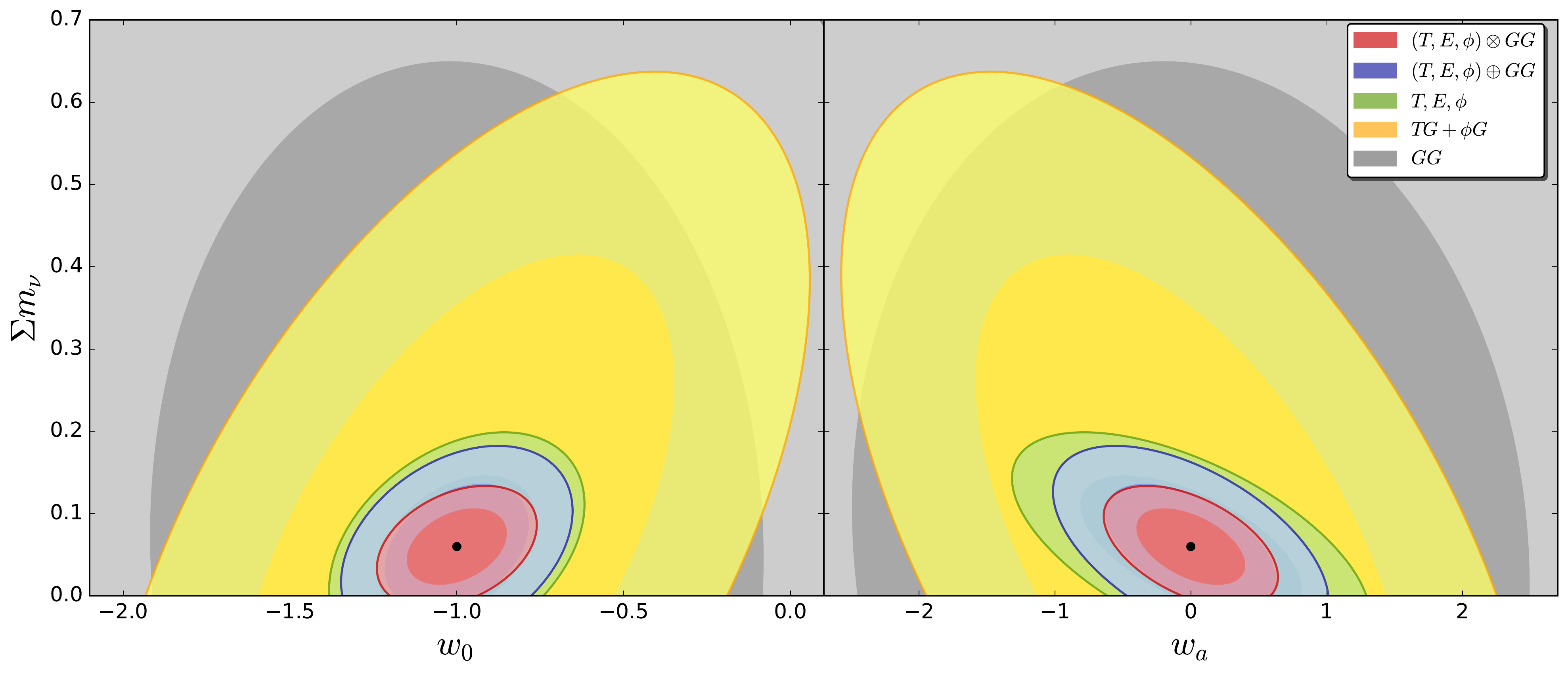}
\caption{Marginalized 68\% and 95\% 2D confidence regions for the joint constraints on $\Sigma m_\nu$ and $w_0$, $w_a$ for a 9 parameters model ($w_0 w_a$CDM+$\Sigma m_\nu$) using the combination of $Planck$+SO and Euclid-ph-like in the top panel and of LiteBIRD+CMB-S4 and SKA1 in the bottom panel. The grey contours correspond LSS-only constraints ($GG$), the yellow contours to the cross-correlation only ($TG+\phi$G) the green contours to the CMB-only constraints ($T,E,\phi$), the blue contours to the combination of CMB and galaxy clustering as uncorrelated and the red ones to the combination including cross-correlation.}
\label{fig:3ellipses}
\end{figure}
For the $\Lambda$CDM model the improvement in the uncertainties in cosmological parameters due to the CMB-GC cross-correlation is maximum for $\Omega_c h^2$, whose uncertainty improves by $\lesssim 20 \%$ in the combinations of $Planck$ with EMU and SKA1. The uncertainties for the various configurations are displayed in Tab.~\ref{tab:LCDM}. Since it has been discussed \cite{1311.0905,1710.09465} that CMB-GC cross-correlation can help to constrain fluctuation amplitudes, we also derive the uncertainty on $\sigma_8$ for the $\Lambda$CDM and the extCDM models using a Jacobian matrix transformation.

For the dark energy extensions, the uncertainties on the parameters of state can be reduced up to a factor $\lesssim$ 2 with the inclusion of CMB-GC cross-correlation. In the simpler $w_0$CDM model, we obtain the best uncertainty on the dark energy parameter of state ($\sigma(w_0) \sim 0.025$)  with the combination of LiteBIRD+S4 $\otimes$ Euclid-ph-like. CMB-GC cross-correlation can complement those surveys that in the uncorrelated CMB $\oplus$ GC combination do not achieve the best constraints: as an example, for the $w_0$CDM model the error on $w_0$ from $Planck$+SO $\oplus$ SKA1 combination is improved by $\sim 40\%$ when adding the cross-correlation. For the $w_0 w_a$ model, the LiteBIRD+S4 $\otimes$ Euclid-ph-like combination provides also the best constraints. In some cases, the dark energy FoM can be improved up to a factor $\lesssim$ 2.5 with the inclusion of CMB-GC cross-correlation.

For the neutrino sector, while $N_{\rm eff}$ is mainly constrained from CMB as already said, the uncertainty on the neutrino mass can be reduced by CMB-GC cross-correlation up to a factor $\sim$ 40\%. We forecast a $\sim$ 2.5$\sigma$ detection ($\sigma(\Sigma m_\nu) \sim 25$ meV) of the neutrino mass from the 2D joint analysis of CMB and GC for the combinations of LiteBIRD+S4 with Euclid-ph-like, LSST and SKA1, and an almost 4$\sigma$ detection ($\sigma(\Sigma m_\nu) \sim 16$ meV) for the combination of LiteBIRD+S4 with SPHEREx.

\begin{figure}
\centering
 \includegraphics[width = \columnwidth]{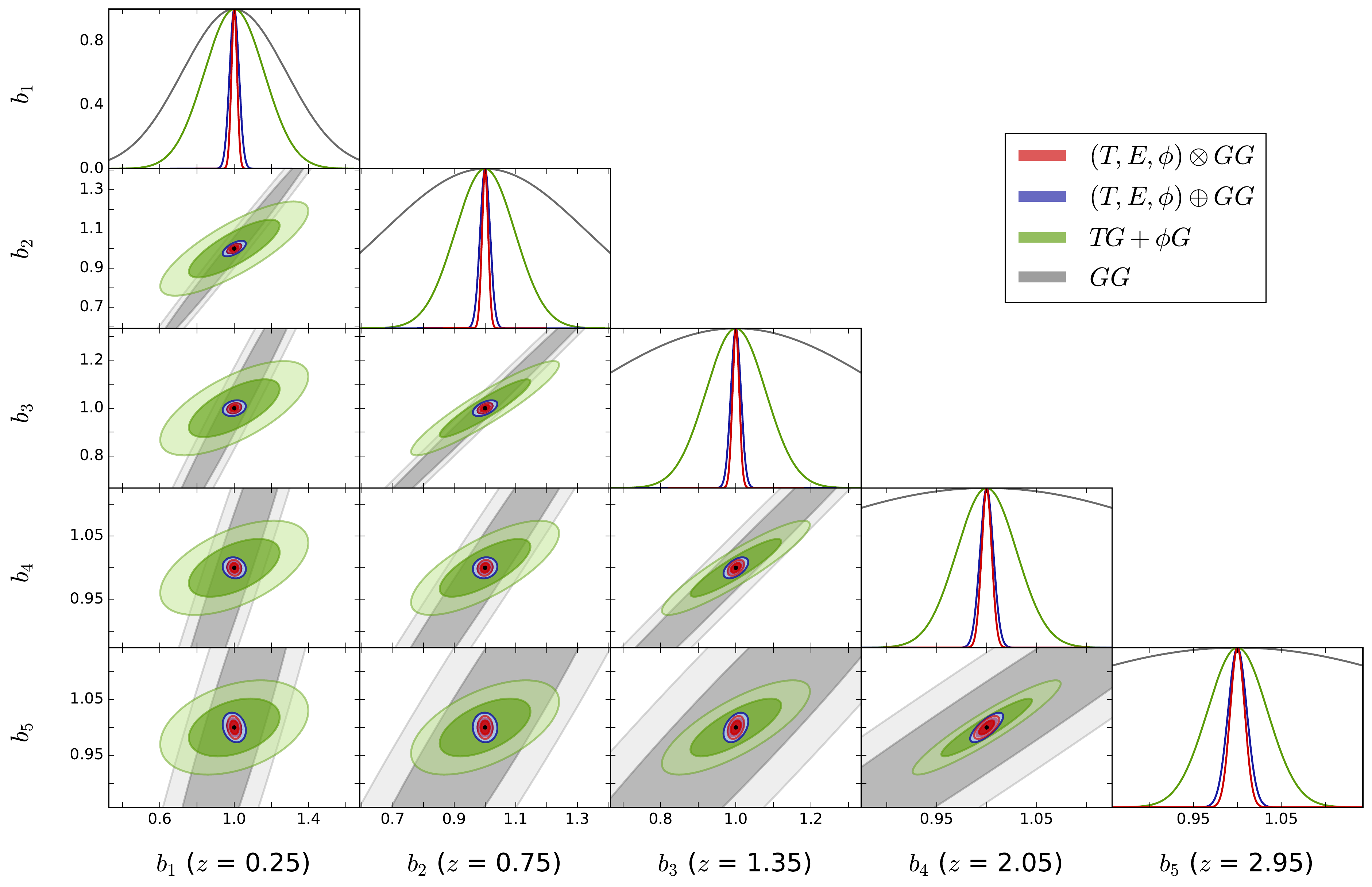}
\caption{Marginalized 68\% and 95\% 2D confidence regions for the joint constraints on the bias nuisance parameters using the combination of LiteBIRD+S4 and SKA1.  The grey contours correspond LSS-only constraints ($GG$), the green contours to the cross-correlation only ($TG+\phi$G), the blue contours to the combination of CMB and galaxy clusering as uncorrelated and the red ones to the combination including cross-correlation.}
\label{fig:bias}
\end{figure}

For the primordial universe sector, since the CMB-GC cross-correlation impact on the running uncertainty is negligible, we recover similar constraints on $\d n_s/ \d \ln k$ to the results from a joint analysis of the CMB and the 3D galaxy power spectra $P(k)$ given in \cite{1606.03747,1611.05883}. For $f_{\rm NL}$, the impact of CMB-GC cross-correlation is maximal when using the single bin configuration since $f_{\rm NL}$ is not constrained from the GC autospectra without tomography, as it is also shown in \cite{1810.06672}. Considering the full CMB $\otimes$ GC tomographic approach, we forecast an uncertainty $\sigma(f_{\rm NL}) \sim 1.5$ for SKA1 and $\sigma(f_{\rm NL}) \sim 2$ for its precursor EMU. Due to their large $f_{\rm sky}$ and redshift depth, we note that radio continuum surveys will perform better than Euclid-like, LSST and SPHEREx for detecting the scale-dependent bias induced by $f_{\rm NL}$. Let us also note that the minimum multipole $\ell_{\rm min}$ is quite critical for $\sigma(f_{\rm NL})$. In order to quantify the effect of this choice, we compute the Fisher matrix for the tomographic combination of $Planck$ and EMU (which has $\ell_{\rm min}$ = 2) using instead $\ell_{\rm min}$ = 20, as for LSST. We obtain $\sigma(f_{\rm NL})$ = 2.8 for $Planck$ $\otimes$ EMU and $\sigma(f_{\rm NL})$ = 6.2 for $Planck$ $\oplus$ EMU, which compared to the errors in Tab.~\ref{tab:primordial} (2.1 and 2.8, respectively) shows a degradation of the constraints, in particular for the galaxy autospectra.

In Fig.~\ref{fig:ellipses} we compare the constraints from the CMB, GC and their cross-correlation with the errors from the CMB $\oplus$ GC and CMB $\otimes$ GC combinations for the three 2 parameters extensions studied (dark energy, neutrino physics and primordial universe). We consider two different cases: $Planck$ $\otimes$ LSST and S4 $\otimes$ SKA1. For the primordial universe extension, we do not show the CMB information since in our analysis $f_{\rm NL}$ is only constrained from the induced scale-dependent bias which enters in the $GG$, $TG$ and $\phi G$ angular power spectra. The CMB-GC cross-correlation information independently is capable for providing competitive constraints on the local non-Gaussianity parameter: for the best case, when cross-correlating LiteBIRD+S4 with SKA1 we obtain $\sigma(f_{\rm NL}) \sim 2.4$ from $TG+\phi G$.

\begin{figure}
\centering
\makebox[\textwidth]{
\includegraphics[width =\columnwidth]{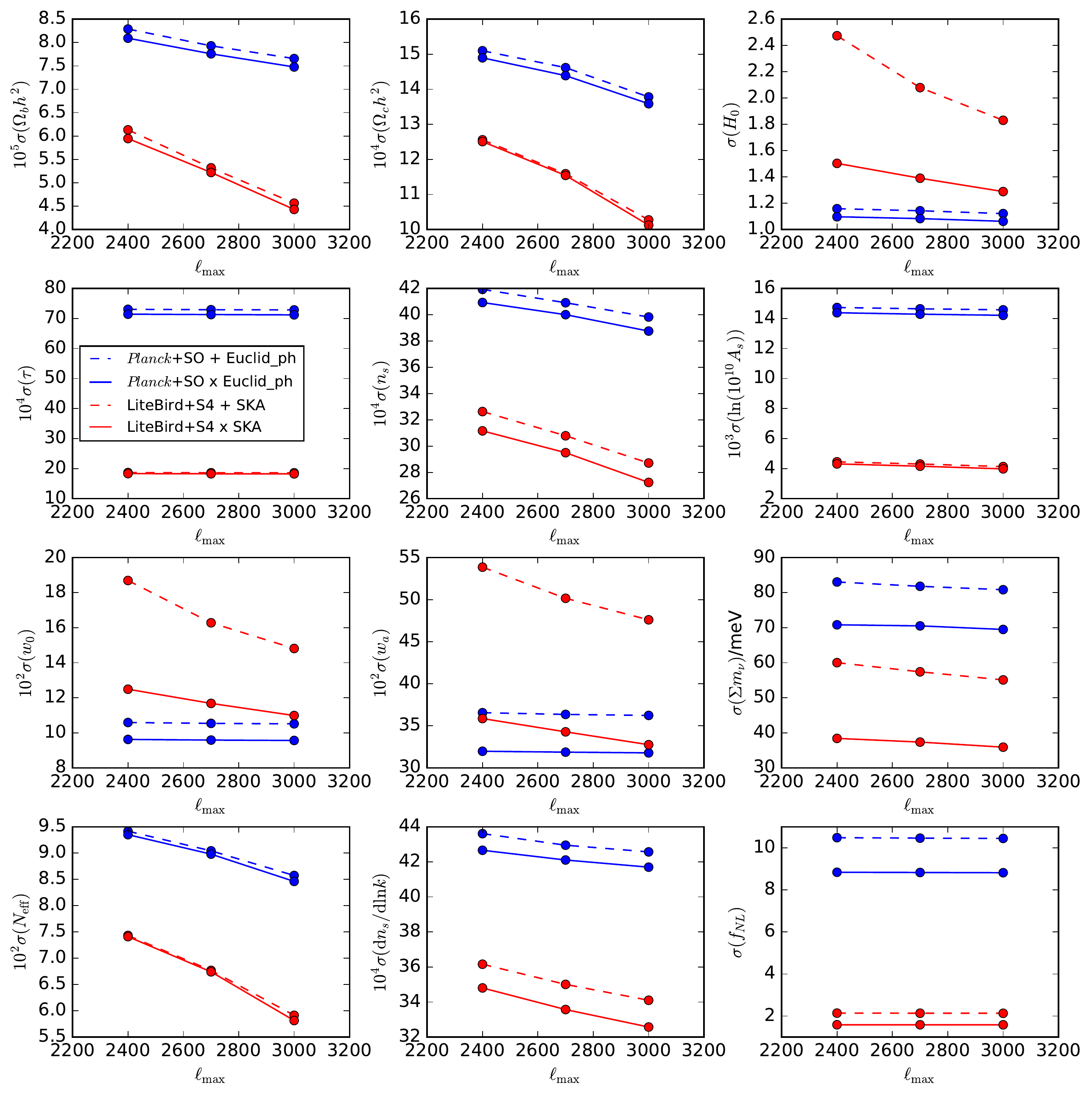}
}
\caption{68\% marginalized constraints on the 12 cosmological parameters of the extCDM model as a function of $\ell_{\rm max}$ of the CMB for the tomographic combinations of $Planck$+SO with Euclid-ph-like and LiteBIRD+S4 with SKA1. The dashed lines correspond to the constraints from CMB $\oplus$ GC and the solid lines to the ones from CMB $\otimes$ GC.}
\label{fig:lmax}
\end{figure}

\begin{figure}
\centering
\makebox[\textwidth]{
\includegraphics[width =\columnwidth]{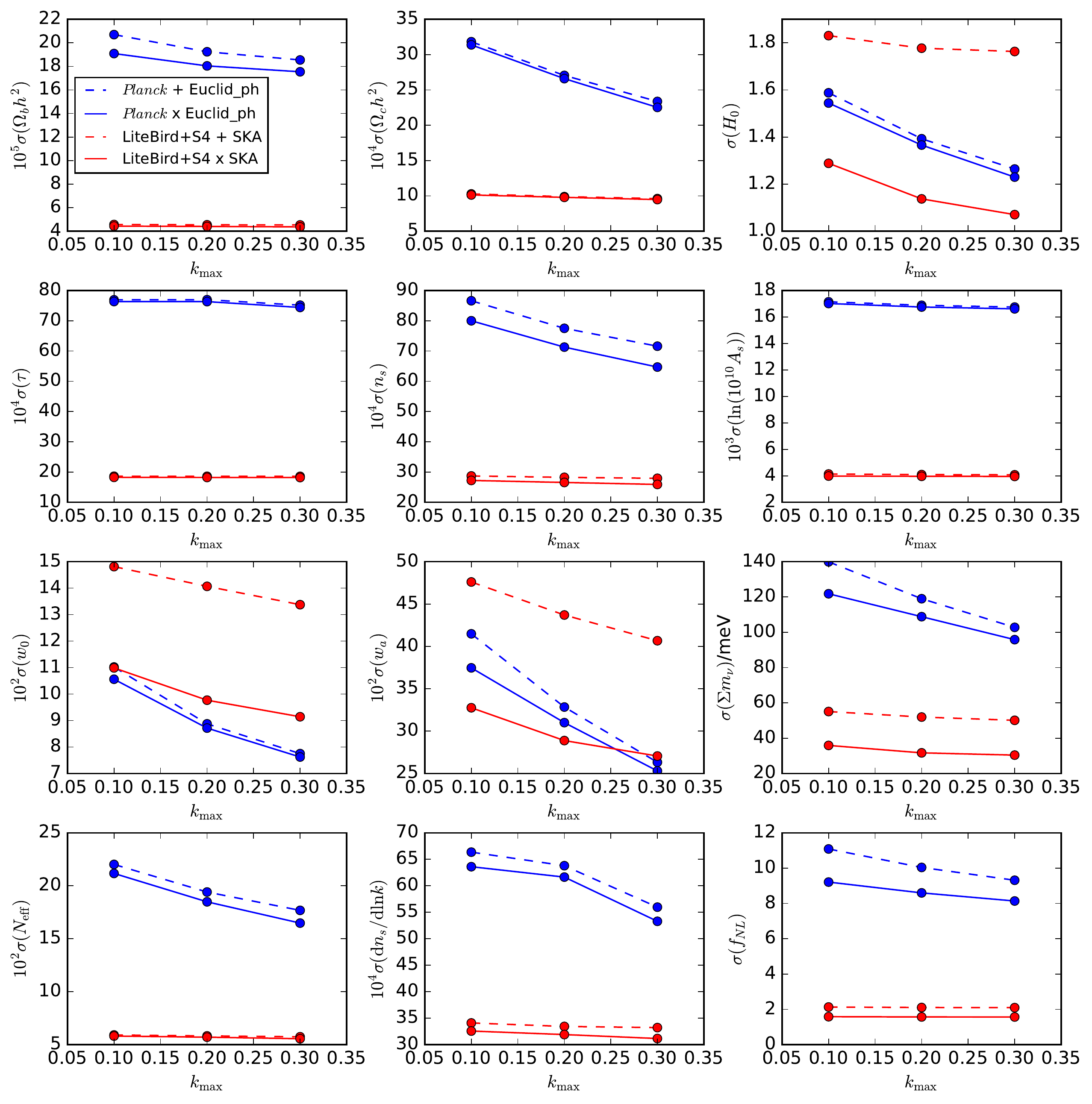}
}
\caption{68\% marginalized constraints on the 12 cosmological parameters of the extCDM model as a function of $k_{\rm max}$ for the tomographic combinations of $Planck$ with Euclid-ph-like and LiteBIRD+S4 with SKA1. The dashed lines correspond to the constraints from CMB $\oplus$ GC and the solid lines to the ones from CMB $\otimes$ GC.}
\label{fig:kmax}
\end{figure}

For the 12 parameters extCDM model, we find that the inclusion of the CMB-GC cross-correlation mainly improves the constraints on $H_0$, the parameters of state of dark energy, the neutrino mass and $f_{\rm NL}$. Parameters like $N_{\rm eff}$ and $\d n_s/ \d \ln k$ are mainly constrained by CMB alone and their uncertainties are only marginally improved by adding galaxy surveys on quasi-linear scales.  We derive as well for this model the uncertainties on $\sigma_8$ and find that CMB-GC cross-correlation can improve up to a factor $\lesssim$ 2 the constraints on this parameter for the combination of $Planck$ with SKA1.

CMB-GC cross-correlation can help in breaking degeneracies which remain in the uncorrelated combination of CMB and GC. 
It has been shown that the neutrino mass limit is model dependent, and in particular it becomes weaker for cosmologies with extended dark energy models and modified gravity. This degeneracy was first noticed in \cite{Hannestad:2005gj} and then observed in real data analysis such as \cite{Vagnozzi:2017ovm,RoyChoudhury:2019hls,Ballardini:2020iws}.
Here, we forecast the uncertainties on the 9 parameters model $w_0 w_a$CDM+$\Sigma m_\nu$ for the combinations of $Planck$+SO with Euclid-ph-like and LiteBIRD+S4 with SKA1. In Fig.~\ref{fig:3ellipses} we show the 68\% and 95\% confidence regions for the $w_0$-$\Sigma m_\nu$ and $w_a$-$\Sigma m_\nu$ planes obtained from the CMB, GC and cross-correlation information independently and from the uncorrelated CMB $\oplus$ GC and full CMB $\otimes$ GC combinations. The orientation of the $TG+\phi G$ ellipses is found to be different with respect to the $GG$ contours, which helps in reducing the joint uncertainties.

CMB-GC cross-correlation has also the capability of constraining the galaxy bias parameters, in particular for the single bin cases since the GC without tomography does not constrain the bias. In the multiple bin cases, CMB-GC cross-correlation can increase the FoM of the bias parameters up to a factor $\lesssim 2$. We show in Fig.~\ref{fig:bias} th e constraints by the various probes on the 5 bias parameters for the tomographic combination LiteBIRD+S4 and SKA1.

As discussed in Section~\ref{sec:data}, our $Planck$-like forecasts are obtained by using $\ell_{\rm max} = 1500$ in order to reproduce the $Planck$ 2018  uncertainties for parameters for the baseline cosmology, whereas the high-$\ell$ likelihood reaches $\ell_{\rm max}=2500 (2000)$ in temperature (polarization) and includes several foregrounds residuals and secondary anisotropies nuisance parameters. 
We therefore study the impact of a 10\% and 20\% reduction $\ell_{\rm max}$ for SO/CMB-S4, which means to adopt $\ell_{\rm max}^{T,E} = 2700$, $\ell_{\rm max}^{\phi} = 900$ and $\ell_{\rm max}^{T,E} = 2400$, $\ell_{\rm max}^{\phi} = 800$, respectively. For these two cases, we compute the constraints on the cosmological parameters for the extCDM model using the tomographic combinations of $Planck$+SO with Euclid-ph-like and LiteBIRD+S4 with SKA1. We represent in Fig.~\ref{fig:lmax} the constraints as a function of the CMB maximum multipole. It is shown that for some parameters like $H_0$ and $w_0$, adopting a more conservative cut in the CMB increases the relative importance of the CMB-GC cross-correlation.

 We also explore the behavior of the constraints when using scales smaller than $k_{\rm max}$ = 0.1 $h$/Mpc in the analysis. We take the tomographic combinations of $Planck$ with Euclid-ph-like and S4 with SKA1, and calculate the uncertainties on the extCDM model from CMB $\oplus$ GC and CMB $\otimes$ GC using $k_{\rm max}$ = 0.2 $h$/Mpc and $k_{\rm max}$ = 0.3 $h$/Mpc. In Fig.~\ref{fig:kmax} we show the errors for the 12 cosmological parameters as function of $k_{\rm max}$. We find improvements with $k_{\rm max}$ for the constraints on majority of the parameters, except for those like $\tau$ that are constrained by CMB or $f_{\rm NL}$ which is mainly constrained from large scales. The neutrino mass and the dark energy parameters of state are those that have benefited most from the increase of $k_{\rm max}$.
 
 Let us finally note that we have considered experimental specifications of CMB instruments which have already operated/or in preparation/or funded. We have not studied in depth the concept for a next CMB space mission dedicated to polarization with an angular resolution would allow a CMB lensing reconstruction much better than $Planck$ even at low multipoles such as CORE \cite{Delabrouille:2017rct, deBernardis:2017ofr} or PICO \cite{Hanany:2019lle}
 or PRISM \cite{Delabrouille:2019thj}. {\sl For the cosmological model with 12 parameters studied here}, we have checked that the uncertainties in cosmological parameters improve when combining a PRISM-like experiment with EMU or SPHEREx compared to the combination with LiteBIRD+S4, albeit the relative importance of CMB-GC cross-correlation does not change much with respect to the cases discussed here.

\section{Conclusions}
\label{sec:conclusions}
We have studied the cross-correlation between the CMB fields, including lensing, and the galaxy clustering in the perspective of future cosmological surveys. We have used a 2D joint tomographic approach in the harmonic domain to determine the cosmological information contained in the CMB-galaxy clustering cross-correlation. We have considered $Planck$-like, SO and S4 for the CMB and Euclid-like photometric and spectroscopic surveys, LSST, SPHEREx, SKA1 and its precursor EMU for galaxy surveys.
By restricting our analysis on quasi-linear scales our main results are the following:

\vspace{.3cm}

$\bullet$ By a signal-to-noise ratio (SNR) analysis, the $TG$ cross-correlation 
between a CMB full sky survey, such as $Planck$ or LiteBIRD, and a wide and deep LSS survey, such as EMU, reaches its maximum, i. e. $\sim 4.3$.
The CMB lensing-galaxy cross-correlation SNR reaches its maximum, $\sim 200$, for the combination of LiteBIRD+S4 and SKA1, which ensures
high fidelity CMB lensing maps on top of the already cited wide and deep galaxy survey.

\vspace{.3cm}

$\bullet$ We find that tomography plays an important role in the SNR, in particular for SPHEREx, EMU and SKA1.
By considering tomography, the SNRs increase to $\sim 5$ and $\sim 240$, for $TG$ with EMU and $\phi G$ with SKA1, respectively.

\vspace{.3cm}

$\bullet$ We have calculated the importance of including the RSD and general relativity contributions in the angular power spectra of the galaxy number counts and their cross-correlation with the CMB in Appendix A. We find that the RSD and velocity contributions are those which have a larger impact in the galaxy number counts autospectra, while the lensing contribution is the most important one for the cross-correlation spectra. At the level of cosmological parameters, we find that the uncertainties do not change by an amount larger than $\sim$ 1-2\% for a 2D joint analysis of CMB and galaxy clustering and $\sim$ 10\% for galaxy clustering only.

\vspace{.3cm}

$\bullet$ We have then evaluated the relevance of the CMB-galaxy clustering cross-correlation in the uncertainties on parameters by using a Fisher matrix approach and reported the results for several different cosmological models in Appendix B. 
In terms of cosmological parameters, the inclusion of CMB-galaxy clustering cross-correlation in a 2D joint analysis will help in constraining the parameter of state of dark energy and its possible redshift dependence improving up to a factor $\sim 2$ the FoM with respect to the combination of CMB and galaxy clustering,  
in detecting the neutrino mass with nearly $4\sigma$ significance and to squeeze the uncertainty in the primordial non-Gaussianity local parameter to $\sigma(f_{\rm NL}) \sim 1.5-2$ only by using 
the effect of the scale-dependent bias on two point statistics. Cross-correlation will be useful for constraining the uncertainties on $\sigma_8$ and on the galaxy clustering bias as well. 

\vspace{.3cm}

$\bullet$ We have also studied possible modifications to the default choices of the range of multipoles
of the CMB surveys and of considering quasi-linear scales only for the galaxy clustering.
As expected, we find that a reduction of the (effective) resolution of a CMB experiment (which could mimic better
more realistic performances when foreground residuals and CMB secondary anisotropies are taken into account)
increases the relative importance of the CMB-galaxy clustering cross-correlation, whereas the inclusion of more and more
non-linear scales decreases it.

\vspace{.3cm}

This work can be naturally extended in many directions, from the inclusion of galaxy shear to a study of the scale dependence in the galaxy bias which in this paper was considered only for local primordial non-Gaussianity.

\section*{Acknowledgments}
We wish to thank J. Asorey, C. Carbone and M. Archidiacono for useful discussions. We acknowledge partial financial contributions from the agreement ASI/INAF n. 2018-23-HH.0 “Attivit\'a scientifica per la missione EUCLID — Fase D” and from Accordo Attuativo n. 2020-9-HH.0 ASI-UniRM2 “Partecipazione italiana alla fase A della missione LiteBIRD". We also acknowledge the support from the Ministero degli Affari Esteri
della Cooperazione Internazionale - Direzione Generale
per la Promozione del Sistema Paese Progetto di Grande Rilevanza ZA18GR02. JRBC acknowledges financial support by a INAF fellowship. MB was supported by the South African Radio Astronomy Observatory, which is a facility of the National Research Foundation, an agency of the Department of Science and Technology. MB and RM was supported by the South African Radio Astronomy Observatory, which is 
a facility of the National Research Foundation, an agency of the Department of Science and Technology. MB is also supported by a Claude Leon Foundation fellowship. RM is also supported by the UK Science \& Technology Facilities Council (Grant ST/N000668/1). DP acknowledges financial support by the ASI Grant 2016-24-H.0. JARM acknowledges financial support from the Spanish Ministry of
Science and Innovation under the project AYA2017-84185-P.
We also acknowledge the support from the Ministero degli Affari Esteri della Cooperazione Internazionale - Direzione Generale per la Promozione del Sistema Paese Progetto di Grande Rilevanza 
ZA18GR02.

\appendix
\section{Angular power spectrum of the total number counts}
\label{sec:relativistic}

In this appendix we define all the RSD and general relativity contributions to the galaxy number counts. As remarked in Section \ref{sec:spectra}, these terms are included in our analysis (see \cite{Challinor:2011bk,1812.01636} for details).

The observed number counts $\Delta_{\ell}(k,z)$ can be split in the density term plus various corrections:
\begin{equation}
    \Delta_{\ell} (k,z) = \Delta^{\rm D}_\ell + \Delta^{\rm RSD}_\ell + \Delta^{\rm L}_\ell + \Delta^{\rm V}_\ell + \Delta^{\rm TD}_\ell + \Delta^{\rm ISW}_\ell+ \Delta^{\rm P}_\ell
\end{equation}

The density term $\Delta^{\rm D}_\ell$ is defined as: 
\begin{equation}
\Delta^{\rm D}_\ell = b(z)\delta^{\rm s}_k(z) j_\ell(k\chi)  + \left(b_e-3\right){\cal H}(z)\frac{v_k}{k} j_\ell(k\chi) 
\end{equation}
where $\delta^{\rm s}_k(z)$ is the synchronous-gauge linear matter density perturbation and $b_e$ is the evolution bias, which is defined as $    b_e = \partial [a^3 \Bar{N}]/\partial \ln{a} $, here $\Bar{N}$ is the number density of background sources.
We can rewrite $\Delta^{\rm D}_\ell = \delta^{\rm n}_k(z) j_\ell(k\chi) $ if we define the Newtonian-gauge density perturbation as:
\begin{equation}
  \delta^{\rm n}_k(z) = b_G(z)\delta^{\rm s}_k(z) + \left(b_e-3\right){\cal H}(z)\frac{v_k}{k}
\end{equation}
where $b_G(z)$ is the galaxy bias and $v_k$ is the Newtonian-gauge velocity of the sources.

\begin{figure}
\centering
\makebox[\textwidth]{
\includegraphics[width = \columnwidth]{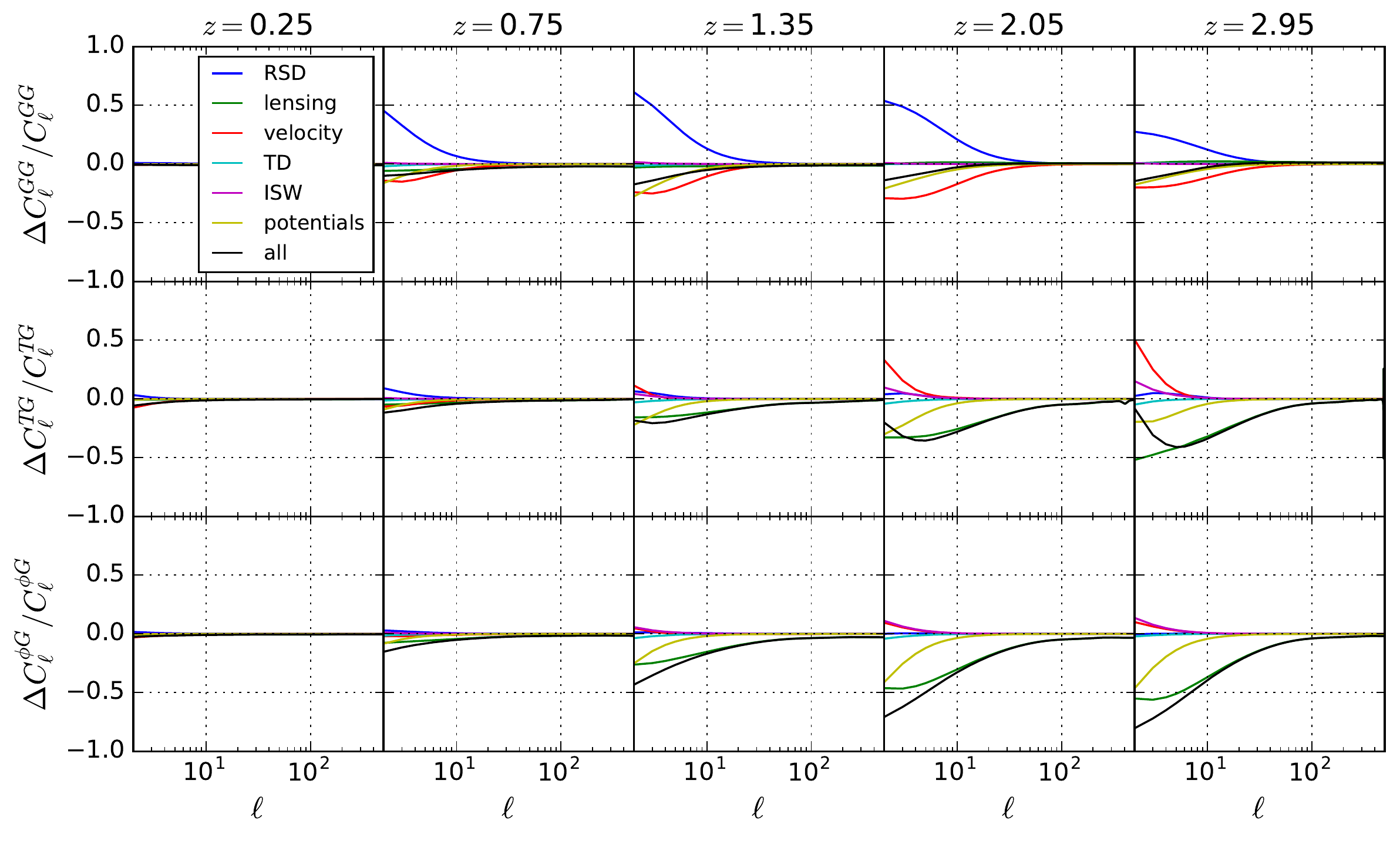}
}
\caption{Relative impact of the contributions to the total number counts with respect to the density-only angular power spectra for the $GG$ autospectra (top panel) and for the $TG$ (mid panel) and $\phi G$ (bottom panel) cross-correlation spectra of the 5 SKA1 redshift bins. The coloured lines correspond to the impact of each single term and the black lines to the sum of all the contributions.}
\label{fig:corrections_ska}
\end{figure}

\begin{figure}
\centering
\makebox[\textwidth]{
\includegraphics[width = \columnwidth]{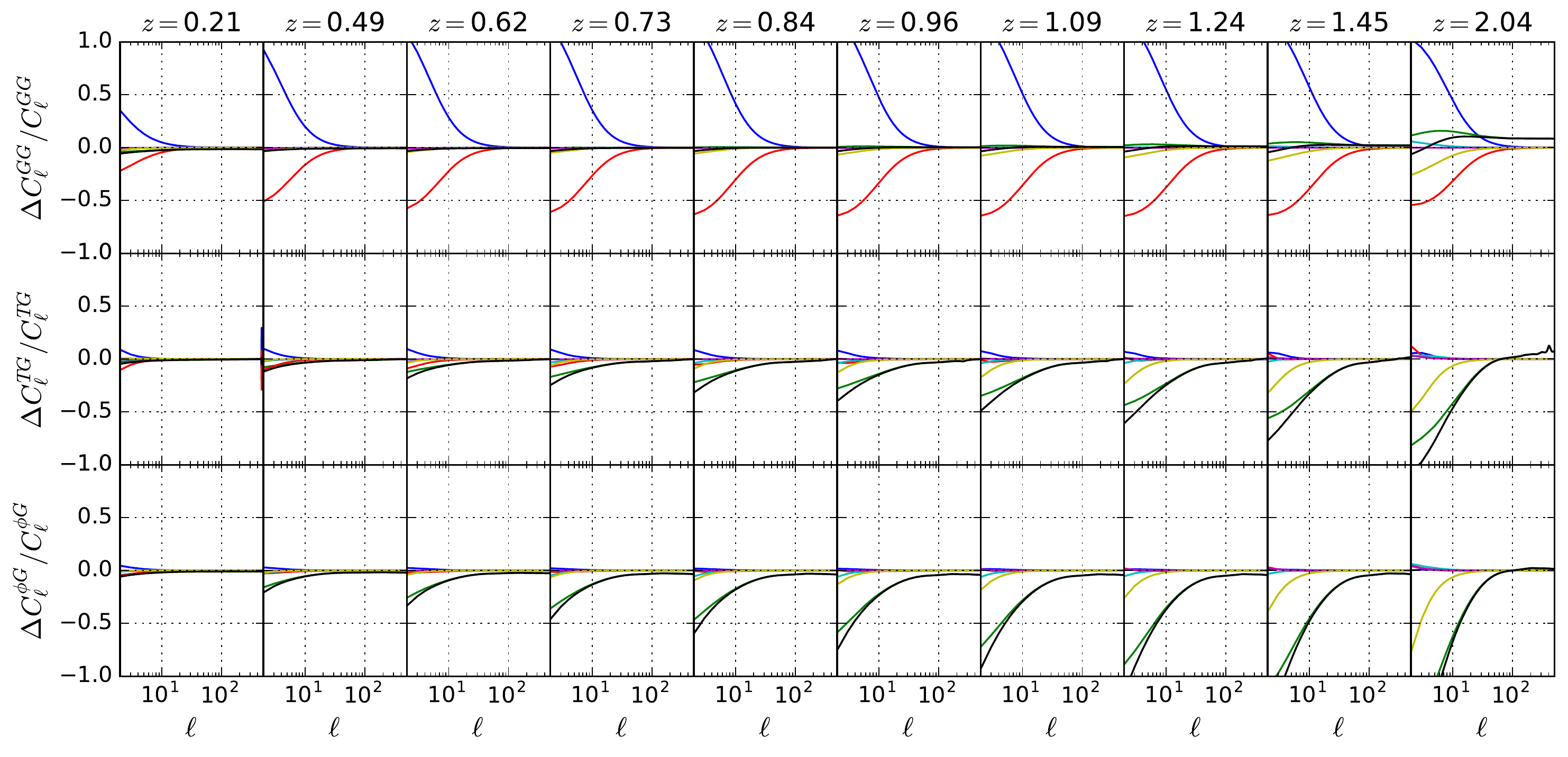}
}
\caption{Relative impact of the contributions to the total number counts with respect to the density-only angular power spectra for the $GG$ autospectra (top panel) and for the $TG$ (mid panel) and $\phi G$ (bottom panel) cross-correlation spectra of the 10 Euclid-ph-like redshift bins. The coloured lines correspond to the impact of each single term and the black lines to the sum of all the contributions. }
\label{fig:corrections_euclid}
\end{figure}

The term $\Delta^{\rm RSD}_\ell$ accounts for the redshift space distorsions:
\begin{equation}
    \Delta^{\rm RSD}_\ell = \frac{k v_k}{\cal H} j_\ell''(k\chi) 
\end{equation}

The lensing term $\Delta^{\rm L}_\ell$ accounts for the contribution from the lensing convergence:
\begin{equation}
  \Delta^{\rm L}_\ell =  \left[\phi_k(\chi)+\psi_k(\chi)\right]j_\ell(k\chi) \frac{\ell(\ell+1)}{2} \int_0^\chi {\rm d}\chi' \frac{\chi-\chi'}{\chi\chi'}(2-5s) 
\end{equation}
where $s \equiv \partial \log \Bar{N_s}/\partial m_{*}$ is the magnification bias. For the functional form of the redshift dependence, $s(z)$, we derive the magnification bias for LSST, EMU and SKA1 based on the luminosity functions and methodology by \cite{1505.07596}. For the Euclid-like surveys, we adopt the $s(z)$ functional forms by \cite{1507.04605} for the photometric sample and by \cite{Bermejo-Climent:2019spz} for the spectroscopic sample. For SPHEREx, since there is not in the literature a derivation of its magnification bias or an estimation for its luminosity function, we assume a constant value $s(z) = 0.42$.

The velocity term $\Delta^{\rm V}_\ell$ accounts for the Doppler effect due to peculiar motions:
\begin{equation}
   \Delta^{\rm V}_\ell = \left[ \frac{2-5s}{{\cal H}(z)\chi} + 5s - b_e + \frac{\dot{\cal H}}{{\cal H}^2} \right]v_k j_\ell'(k\chi) 
\end{equation}    

We consider also the ISW term ($\Delta^{\rm ISW}_\ell$), the time-delay term ($\Delta^{\rm TD}_\ell$) and other local contributions from gravitational potentials ($\Delta^{\rm P}_\ell$).

\begin{equation}
    \Delta^{\rm ISW}_\ell = \left[\dot\phi_k(\chi)+\dot \psi_k(\chi) \right] j_{\ell}(k\chi) 
    \int_0^\chi {\rm d} \chi' \left[ \frac{2-5s}{{\cal H}\chi'} + 5s - b_e + \frac{\dot{\cal H}}{{\cal H}^2} \right] 
\end{equation}

\begin{equation}
 \Delta^{\rm TD}_\ell =\left[\phi_k(\chi)+\psi_k(\chi)\right] j_\ell(k\chi)  
\int_0^{\chi} {\rm d}\chi' \frac{2-5s}{\chi'}    
\end{equation}

\begin{equation}
    \Delta^{\rm P}_\ell = \left\{\left[ \frac{2-5s}{{\cal H}(z)\chi} + 5s - b_e + \frac{\dot{\cal H}}{{\cal H}^2} + 1\right]
\psi_k + (5s-2)\phi_k + \frac{\dot{\phi}_k}{\cal H}(z) \right\} j_\ell(k\chi) 
\end{equation}

We now quantify the impact of the contributions defined here on the $GG$ autospectra and the $TG$, $\phi G$ cross-correlation spectra for the redshift bins of SKA1 and Euclid-ph-like.
In Figs.~\ref{fig:corrections_ska} and \ref{fig:corrections_euclid} we show the relative differences on the spectra obtained for each term, together and independently, with respect to the spectra including the density term only. We obtain that the relative weight of the corrections is larger at higher $z$ and large scales ($\ell \lesssim 100$). The RSD and velocity terms are the most important for the $GG$ spectra, while the lensing contribution has a larger impact on the cross-correlation spectra.

We also check whether neglecting the RSD and general relativity corrections could have an impact on the cosmological parameter uncertainties. For this, we compute the Fisher matrix for the combination of $Planck$+SO with the Euclid-ph-like and SKA1 datasets neglecting all the contributions to the angular power spectra beyond the density term. We find that once combined with the CMB the constraints on the parameters do not change by a larger amount than $\sim$ 1-2\% with respect to the case including all the contributions. Instead, when considering the constraints by the galaxy number counts alone the errors on the cosmological parameters can vary around $\sim$ 10\%.

\section{Forecast constraints on cosmological parameters}
\label{sec:constraints}

In this appendix we present the 68\% marginalized constraints on the parameters for the various cosmological models discussed in Section \ref{sec:results}. 
We present the $\Lambda$CDM model in Tab.~ \ref{tab:LCDM}, the $w_0$CDM model in Tab.~ \ref{tab:w0CDM}, 
the 2 parameters extensions of $\Lambda$CDM (dynamical dark energy, neutrino physics and primordial universe) in Tabs.~\ref{tab:w0waCDM}-\ref{tab:primordial}, and the 12 parameters extCDM model in Tab.~\ref{tab:extCDM}. All the errors are marginalized over the bias and nuisance parameters and correspond to the analysis up to quasi-linear scales with $k_{\rm max}$ = 0.1 $h$/Mpc.

\begin{table}
\scriptsize
\caption{68\% marginalized constraints on the parameters of the $\Lambda$CDM cosmology obtained from the combination of each pair of CMB and galaxy surveys, in the single bin and tomographic configurations. We show the constraints from the full combination including the CMB-GC cross-correlation (CMB $\otimes$ GC) and, between parenthesis, the constraints from the uncorrelated combination of Fisher matrices (CMB $\oplus$ LSS). We also list FoM of the 6 cosmological parameters. For each survey combination, we mark in boldface the cosmological parameter which benefits most from including the CMB-LSS cross-correlation.}

\resizebox{\textwidth}{!}{
\begin{tabular}{|l|l|cccccccccccc|} 
\hline
\multicolumn{2}{|c}{$\Lambda$CDM} &  
\multicolumn{2}{|c}{Euclid-ph-like} &  \multicolumn{2}{c}{Euclid-sp-like} &
\multicolumn{2}{c}{LSST} &  \multicolumn{2}{c}{SPHEREx} &
\multicolumn{2}{c}{EMU} & \multicolumn{2}{c|}{SKA1}\\
\hline
\multirow{1}{*}{CMB} & \multirow{2}{*}{Parameter} & 1 bin & 10 bins & 1 bin & 9 bins & 1 bin & 10 bins & 1 bin & 10 bins & 1 bin & 5 bins & 1 bin & 5 bins \\

 \multirow{1}{*}{survey}&  &  $\otimes$($\oplus$) & $\otimes$($\oplus$) & $\otimes$($\oplus$) & $\otimes$($\oplus$) & 
 $\otimes$($\oplus$) & $\otimes$($\oplus$) & $\otimes$($\oplus$) & $\otimes$($\oplus$) &
 $\otimes$($\oplus$)& $\otimes$($\oplus$) & $\otimes$($\oplus$) & $\otimes$($\oplus$) 
\\
\hline 
\multirow{9}{*}{$Planck$} & $10^5$ $\sigma(\Omega_b h^2)$ & 14(14) & 11(11) & 14(14) & 12(12) & 14(14) & 12(12) & 14(14) & 12(11) & 14(14) & 12(13) & 14(14) & 12(12)  \\

& $10^4$ $\sigma(\Omega_c h^2)$
& 12(12) & \bf{5.8(6.2)} & 12(12) & 8.5(8.5) & 11(11) & \bf{6.8(7.7)} & 11(11) & \bf{6.3(7.1)} & 11(11) & \bf{8.2(9.9)} & 11(11) & \bf{7.5(9.0)} \\

& $10^2$ $\sigma(H_0)$ & \bf{56(57)} & 26(27) & \bf{56(57)} & 38(38) & 51(52) & 31(34) & 52(52) & 28(32) &  \bf{51(52)} & 37(44) & 51(52) & 34(40)\\

& $10^4$ $\sigma(\tau)$ & 74(74) & 66(67) & 74(74) & 69(69) & 72(72) & 67(68) & 72(72) & 67(67) & 72(72) & 69(70) & 72(72) & 67(68)  \\

& $10^4$ $\sigma(n_s)$ & 36(36) & 29(29) & 36(36) & 32(32) & 34(35) & 30(31) & 35(35) & 30(30) & 35(35) & 31(33) & \bf{34(35)} & 31(33)  \\

& $10^3$ $\sigma(\ln (10^{10} A_s))$ & 14(14) & 13(13) & 14(14) & 13(13) & 13(13) & 13(13) & 13(13) & 13(13) & 13(13) & 13(13) & 13(13) & 12(13)   \\

& $10^4$ $\sigma(\sigma_8)$ & 62(63) & 50(53) & 62(63) & 54(54) & \bf{53(55)} & 50(53) & 55(55) & 51(53) & 55(55) & 48(54) & 54(55) & 46(52)  \\

& FoM$_{\rm cosmo}$ / $10^4$ & 15(15) & 21(20) & 15(15) & 18(18) & 16(16) & 20(19) & 16(16) & 20(19) & 16(16) & 19(17) & 16(16) & 20(18)\\

& FoM$_{\rm bias}$ / $10^2$ & 0.21(0.03) & 645(619) & 0.020(0.005) & 35(35) & 0.69(0.09) & 68(57) & \bf{0.026(0.001)} & 47(45) & 12(3.4) & 117(109) & 1.5(0.8) & 135(126)  \\

\hline 
\multirow{9}{*}{$Planck$+SO} & $10^5$ $\sigma(\Omega_b h^2)$ & 4.7(4.7) & 4.5(4.5) & 4.7(4.7) & 4.6(4.6) & 4.7(4.7) & 4.5(4.5) & 4.7(4.7) & 4.5(4.5) & 4.7(4.7) & 4.6(4.6) & 4.7(4.7) & 4.6(4.6) \\

& $10^4$ $\sigma(\Omega_c h^2)$
& 6.2(6.2) & \bf{4.6(4.9)} & 6.2(6.2) & 5.7(5.7) & 6.2(6.2) & \bf{4.8(5.5)} & 6.2(6.2) & \bf{4.7(5.2)} &6.2(6.2) & \bf{5.6(6.0)} & 6.2(6.2) & \bf{5.1(5.8)}  \\

&$10^2$ $\sigma(H_0)$ & 24(24) & 17(18) & 24(24) & 21(21) & 24(24) & 18(21) & 24(24) & 18(20) & 24(24) & 22(23) & 24(24) & 20(22) \\

& $10^4$ $\sigma(\tau)$ & 56(56) & 48(49) & 56(56) & 53(53) & 56(56) & 49(52) & 56(56) & 49(51) & 56(56) & 53(55) & 56(56) & 51(54) \\

& $10^4$ $\sigma(n_s)$ & 19(19) & 19(19) & 19(19) & 19(19) & 19(19) & 19(19) & 19(19) & 19(19) & 19(19) & 19(19) & 19(19) & 19(19) \\

& $10^3$ $\sigma(\ln (10^{10} A_s))$ & 10(10) & 8.5(9.0) & 10(10) & 9.6(9.6) & 10(10) & 8.7(9.5) & 10(10) & 8.8(9.2) & 10(10) & 9.5(9.9) & 10(10) & 9.0(9.6)  \\

& $10^4$ $\sigma(\sigma_8)$ & \bf{30(31)} & 29(30) & 31(31) & \bf{30(31)} & \bf{30(31)} & 28(30) & 31(31) & 30(30) & 31(31) & 30(31) & \bf{30(31)} & 29(30) \\

& FoM$_{\rm cosmo}$ / $10^4$ & 77(77) & 86(84) & 77(77) & 80(80) & 77(77) & 86(81) & 77(77) & 83(82) & 77(77) & 79(78) & 77(77) & 83(80) \\

& FoM$_{\rm bias}$ / $10^2$ & 0.85(0.03) & 760(674) & \bf{0.051(0.005)} & 38(37) & 2.27(0.07) & 94(61) & \bf{0.041(0.001)} & 52(48) & \bf{24(4)} & 146(119) & 4.3(1.0) & 177(142)   \\

\hline

\multirow{9}{*}{LiteBIRD+S4} & $10^5$ $\sigma(\Omega_b h^2)$ & 2.9(2.9) & 2.9(2.9) & 2.9(2.9) & 2.9(2.9) & 2.9(2.9) & 2.9(2.9) & 2.9(2.9) & 2.9(2.9) & 2.9(2.9) & 2.9(2.9) & 2.9(2.9) & 2.9(2.9) \\

& $10^4$ $\sigma(\Omega_c h^2)$
& 2.4(2.4) & 2.2(2.3) & 2.4(2.4) & 2.3(2.3) & 2.4(2.4) & 2.1(2.3) & 2.4(2.4) & 2.2(2.3) & 2.4(2.4) & \bf{2.3(2.4)} & 2.4(2.4) & 2.2(2.3)  \\

& $10^2$ $\sigma(H_0)$ & 9.1(9.2) & 8.3(8.7) & 9.2(9.2) & 9.0(9.0) & 9.1(9.2) & \bf{8.2(8.9)} & 9.2(9.2) & \bf{8.5(8.8)} & 9.2(9.2) & 8.9(9.1) & 9.1(9.2) & \bf{8.4(9.0)}  \\

& $10^4$ $\sigma(\tau)$ & 18(18) & 17(17) & 18(18) & 17(17) & 18(18) & 17(17) & 18(18) & 17(17) & 18(18) & 18(18) & 18(18) & 17(17) \\

& $10^4$ $\sigma(n_s)$ & 13(13) & 13(13) & 13(13) & 13(13) & 13(13) & 13(13) & 13(13) & 13(13) & 13(13) & 13(13) & 13(13) & 13(13) \\

& $10^3$ $\sigma(\ln (10^{10} A_s))$ & \bf{3.2(3.3)} & 3.1(3.2) & 3.3(3.3) & 3.2(3.2) & \bf{3.2(3.3)} & 3.0(3.2) & 3.3(3.3) & 3.1(3.2) & 3.2(3.3) & 3.2(3.2) & \bf{3.2(3.3)} & 3.1(3.2) \\

& $10^4$ $\sigma(\sigma_8)$ & 10(10) & 10(10) & 10(10) & 10(10) & 10(10) & 10(10) & 10(10) & 10(10) & 10(10) & 10(10) & 10(10) & 10(10)\\
& FoM$_{\rm cosmo}$ / $10^4$ & 242(242) & 252(247) & 242(242) & 244(244) & 242(242) & 257(245) & 242(242) & 245(245) & 242(242) & 242(242) & 242(242) & 250(244)  \\

& FoM$_{\rm bias}$ / $10^2$ & 1.85(0.03) & 967(753) & \bf{0.086(0.005)} & \bf{44(42)} & 5.39(0.07) & 138(67) & \bf{0.054(0.001)} & 59(53) & \bf{102(7)} & 876(493) & 24(1.3) & 273(167)    \\

\hline 

\end{tabular}}
\label{tab:LCDM}
\end{table}

 \begin{table}
\scriptsize
\caption{68\% marginalized constraints on $w_0$ for the $w_0$CDM cosmology obtained from the combination of each pair of CMB and galaxy surveys, in the single bin and tomographic configurations. We show the constraints from the full combination including the CMB-GC cross-correlation (CMB $\otimes$ GC) and, between parenthesis, the constraints from the combination of Fisher matrices as uncorrelated (CMB $\oplus$ GC). For each CMB survey, we mark in boldface the galaxy survey for which the error in $w_0$ has a larger improvement by the inclusion of cross-correlation.}
\resizebox{\textwidth}{!}{
\begin{tabular}{|l|c|cccccccccccc|} 
\hline
\multicolumn{2}{|c}{$w_0$CDM} &  
\multicolumn{2}{|c}{Euclid-ph-like} &  \multicolumn{2}{c}{Euclid-sp-like} &
\multicolumn{2}{c}{LSST} &  \multicolumn{2}{c}{SPHEREx} &
\multicolumn{2}{c}{EMU} & \multicolumn{2}{c|}{SKA1}\\
\hline
\multirow{2}{*}{CMB survey} & \multirow{2}{*}{Parameter} & 1 bin & 10 bins & 1 bin & 9 bins & 1 bin & 10 bins & 1 bin & 10 bins & 1 bin & 5 bins & 1 bin & 5 bins \\

 &  &  $\otimes$($\oplus$) & $\otimes$($\oplus$) & $\otimes$($\oplus$) &
 $\otimes$($\oplus$) & $\otimes$($\oplus$) & $\otimes$($\oplus$) & $\otimes$($\oplus$) & $\otimes$($\oplus$) & $\otimes$($\oplus$)& $\otimes$($\oplus$) & $\otimes$($\oplus$) & $\otimes$($\oplus$) 
\\
\hline 
\multirow{1}{*}{$Planck$} & $10^2$ $\sigma(w_0)$& 27(31) & 4.0(4.4)  & 25(31) & 12(12) & 24(28) & 7.1(8.4) & 27(28) & 5.3(5.8) & 22(28) & \bf{8.5(15)}  & 16(27) & 8.2(14)  \\ \hline 
\multirow{1}{*}{$Planck$+SO} & $10^2$ $\sigma(w_0)$& 11(11) & 3.3(3.6) & 11(11) & 7.9(8.2) &  11(11) & 5.0(6.6) & 11(11) & 4.0(4.3) & 11(11) & 6.2(9.4)  & 9.2(11) & \bf{5.3(9.0)}  \\ \hline 
\multirow{1}{*}{LiteBIRD+S4} & $10^2$ $\sigma(w_0)$& 6.1(6.1) & 2.5(2.8) &  6.1(6.1) & 5.2(5.2) & 6.0(6.1) & 3.3(4.6) & 6.1(6.1) & 2.8(3.2) & 6.1(6.1) & 4.2(5.7)  & 5.4(6.1) & \bf{3.6(5.5)}  \\ \hline 

\end{tabular}}
\label{tab:w0CDM}

\bigskip\bigskip

\scriptsize
\caption{68\% marginalized constraints on $w_0$ and $w_a$ for the $w_0 w_a$CDM cosmology obtained from the combination of each pair of CMB and galaxy surveys, in the single bin and tomographic configurations. We show the constraints from the full combination including the CMB-GC cross-correlation (CMB $\otimes$ GC) and, between parenthesis, the constraints from the combination of Fisher matrices as uncorrelated (CMB $\oplus$ GC). We also list the FoM of the two extra parameters. For each survey combination, we mark in boldface the cosmological parameter which benefits most from including the CMB-LSS cross-correlation. We mark as well the FoM which is most improved by cross-correlation.}
\resizebox{\textwidth}{!}{
\begin{tabular}{|l|c|cccccccccccc|} 
\hline
\multicolumn{2}{|c}{$w_0 w_a$CDM} &  
\multicolumn{2}{|c}{Euclid-ph-like} &  \multicolumn{2}{c}{Euclid-sp-like} &
\multicolumn{2}{c}{LSST} &  \multicolumn{2}{c}{SPHEREx} &
\multicolumn{2}{c}{EMU} & \multicolumn{2}{c|}{SKA1}\\
\hline
\multirow{2}{*}{CMB survey} & \multirow{2}{*}{Parameter} & 1 bin & 10 bins & 1 bin & 9 bins & 1 bin & 10 bins & 1 bin & 10 bins & 1 bin & 5 bins & 1 bin & 5 bins \\

 &  &  $\otimes$($\oplus$) & $\otimes$($\oplus$) & $\otimes$($\oplus$) &
 $\otimes$($\oplus$) & $\otimes$($\oplus$)& $\otimes$($\oplus$) & $\otimes$($\oplus$) &$\otimes$($\oplus$) & $\otimes$($\oplus$)& $\otimes$($\oplus$) & $\otimes$($\oplus$) & $\otimes$($\oplus$) 
\\
\hline 
\multirow{3}{*}{$Planck$} & $10^2$ $\sigma(w_0)$& \bf{72(92)} & \bf{10(11)} & \bf{84(92)} & \bf{69(77)} & \bf{75(91)} & \bf{30(37)} & 84(92) & \bf{18(20)} & \bf{74(89)}  & \bf{32(46)} & \bf{63(77)} & \bf{27(47)}  \\

& $10^2$ $\sigma(w_a)$& 166(197) & 32(34) & 188(197) & 149(160) & 169(194) & 72(86) & \bf{176(195)} & 38(41) & 168(190) & 80(106)  & 155(175) & 67(105)  \\

& FoM & 2.2(1.6) & 79(67) & 2.2(1.7) & 5.7(5.2) & 2.4(1.8) & 20(14) & 2.1(1.8) & 50(42) & 2.7(1.9) & 15(6.4)  & 4.0(2.1) & \bf{18(6.9)}  \\
\hline 
\multirow{3}{*}{$Planck$+SO} & $10^2$ $\sigma(w_0)$& 25(26) & \bf{9.4(10)} & 26(26) & \bf{24(25)} &  25(26) & \bf{17(22)} & 26(26) & \bf{14(16)} & 26(26) & 20(23)  & \bf{25(26)} & \bf{15(23)}  \\

& $10^2$ $\sigma(w_a)$& \bf{68(71)} & 28(32) & 71(71) & 58(59) & \bf{68(71)} & 42(53) & \bf{70(71)} & 30(33) & \bf{70(71)} & \bf{53(61)}  & 70(71) & 42(59)  \\

& FoM & 13(13) & 108(89) & 13(13)  & 22(21) & 13(13) & 47(28) & 13(13) & 82(69) & 13(13) & 30(17)  & 16(13) & 45(19)  \\

\hline 
\multirow{3}{*}{LiteBIRD+S4} & $10^2$ $\sigma(w_0)$& 14(14) & 7.7(8.7) & 14(14) & 13(13) & 14(14) & \bf{10(13)} & 14(14) & 10(11) & 14(14)  & \bf{12(14)} & 14(14) & 10(13) \\

& $10^2$ $\sigma(w_a)$& \bf{41(43)} & \bf{22(26)} & 43(43) & 35(35) & 40(43) & 26(33) &  \bf{42(43)} & \bf{22(25)} & \bf{41(43)} & 34(39)  & \bf{41(43)} & \bf{27(37)}  \\

& FoM & 40(38) & 183(140) & 38(38) & 56(54) & 42(38) & 118(65) & 39(39) & 162(126) & 40(38) & 70(45)  & 45(39) & 104(49)  \\

\hline 

\end{tabular}}
\label{tab:w0waCDM}
\end{table}

 \begin{table}
\scriptsize
\caption{68\% marginalized constraints on $\Sigma m_\nu$ and $N_{\rm eff}$ for the $\Lambda$CDM+$\{\Sigma m_\nu,N_{\rm eff} \}$ cosmology obtained from the combination of each pair of CMB and galaxy surveys, in the single bin and tomographic configurations. We show the constraints from the full combination including the CMB-GC cross-correlation (CMB $\otimes$ GC) and, between parenthesis, the constraints from the combination of Fisher matrices as uncorrelated (CMB $\oplus$ GC). We also list the FoM of the two extra parameters. For each survey combination, we mark in boldface the cosmological parameter which benefits most from including the CMB-LSS cross-correlation.}
\resizebox{\textwidth}{!}{
\begin{tabular}{|l|c|cccccccccccc|} 
\hline
\multicolumn{2}{|c}{$\Lambda$CDM+$\{\Sigma m_\nu,N_{\rm eff}\}$} &  
\multicolumn{2}{|c}{Euclid-ph-like} &  \multicolumn{2}{c}{Euclid-sp-like} &
\multicolumn{2}{c}{LSST} &  \multicolumn{2}{c}{SPHEREx} &
\multicolumn{2}{c}{EMU} & \multicolumn{2}{c|}{SKA1}\\
\hline
\multirow{2}{*}{CMB survey} & \multirow{2}{*}{Parameter} & 1 bin & 10 bins & 1 bin & 9 bins& 1 bin & 10 bins & 1 bin & 10 bins & 1 bin & 5 bins & 1 bin & 5 bins \\

 &  &  $\otimes$($\oplus$) & $\otimes$($\oplus$) & $\otimes$($\oplus$) & 
 $\otimes$($\oplus$) & $\otimes$($\oplus$)& $\otimes$($\oplus$) & $\otimes$($\oplus$) &
 $\otimes$($\oplus$) & $\otimes$($\oplus$)& $\otimes$($\oplus$) & $\otimes$($\oplus$) & $\otimes$($\oplus$) 
\\
\hline 
\multirow{3}{*}{$Planck$} & $\sigma(\Sigma m_\nu)$/meV  & \bf{163(173)} & \bf{68(73)} & \bf{166(171)} & 103(103) & \bf{143(149)} & \bf{80(88)} & 149(149) & \bf{55(61)} & \bf{140(149)}  & \bf{101(132)} & \bf{135(149)} & \bf{85(107)}  \\

& $10^2$ $\sigma(N_{\rm eff})$& 22(22) & 18(18) & 22(22) & 20(20) &  22(22) & 19(19) & 22(22) & 20(20) & 21(22) & 20(21)  & 21(22) & 19(20)  \\

& FoM & 28(26) & 85(77) & 28(27) & 47(47) & 33(31) & 67(59) & 31(31) & 92(82) & 34(31) & 50(36)  & 35(31) & 62(47)  \\
\hline 

\multirow{3}{*}{$Planck$+SO} &  $\sigma(\Sigma m_\nu)$/meV & 74(74) & \bf{48(50)} & 74(74) & 65(65) &  74(74) & \bf{51(60)} & 74(74) & \bf{31(43)} & 74(74) & \bf{62(72)}  & 74(74) & \bf{55(68)}  \\

& $10^2$ $\sigma(N_{\rm eff})$& 6.3(6.3) & 6.2(6.2) & 6.3(6.3) & 6.2(6.2) &  6.3(6.3) & 6.2(6.2) & 6.3(6.3) & 6.2(6.2) & 6.3(6.3) & 6.3(6.3)  & 6.3(6.3) & 6.2(6.2)  \\

& FoM & 215(215) & 347(325) & 216(216) &  247(247) & 215(215) & 320(271) & 215(215) & 518(376)  & 215(215) & 258(223)  & 215(215) & 298(238)  \\
\hline

\multirow{3}{*}{LiteBIRD+S4} &  $\sigma(\Sigma m_\nu)$/meV & 40(40) & \bf{24(28)} & 40(40) & 36(36) & 40(40) & \bf{25(34)} & 40(40) & \bf{16(26)} & 40(40)  & \bf{33(39)} & 40(40) & \bf{27(37)} \\

& $10^2$ $\sigma(N_{\rm eff})$&4.0(4.0) & 4.0(4.0) & 4.0(4.0) & 4.0(4.0) &  4.0(4.0) & 3.9(4.0) &  4.0(4.0) & 4.0(4.0) & 4.0(4.0) &4.0(4.0)  & 4.0(4.0) & 4.0(4.0)  \\

& FoM & 618(618) & 1054(893) & 619(619) & 693(693) & 619(619) & 1047(744) & 620(620) & 1525(963) & 619(619) & 751(639) & 619(619)  & 944(674)  \\
\hline

\end{tabular}}
\label{tab:neutrino}

\bigskip
\scriptsize
\caption{68\% marginalized constraints on $\d n_s/ \d \ln k$ and $f_{\rm NL}$ for the $\Lambda$CDM+$\{\d n_s/ \d \ln k$,$f_{\rm NL}\}$ cosmology obtained from the combination of each pair of CMB and galaxy surveys, in the single bin and tomographic configurations. We show the constraints from the full combination including the CMB-GC cross-correlation (CMB $\otimes$ GC) and, between parenthesis, the constraints from the combination of Fisher matrices as uncorrelated (CMB $\oplus$ GC). We also list the FoM of the two extra parameters. For each survey combination, we mark in boldface the cosmological parameter which benefits most from including the CMB-LSS cross-correlation.}
\resizebox{\textwidth}{!}{
\begin{tabular}{|l|c|cccccccccccc|} 
\hline
\multicolumn{2}{|c}{$\Lambda$CDM+$\{\d n_s/ \d \ln k$,$f_{\rm NL}\}$} &  
\multicolumn{2}{|c}{Euclid-ph-like} &  \multicolumn{2}{c}{Euclid-sp-like} &\multicolumn{2}{c}{LSST} &  \multicolumn{2}{c}{SPHEREx} &
\multicolumn{2}{c}{EMU} & \multicolumn{2}{c|}{SKA1}\\
\hline
\multirow{2}{*}{CMB survey} & \multirow{2}{*}{Parameter} & 1 bin & 10 bins  & 1 bin & 9 bins   & 1 bin & 10 bins  & 1 bin & 10 bins & 1 bin & 5 bins & 1 bin & 5 bins \\

 &  &  $\otimes$($\oplus$) & 
 $\otimes$($\oplus$) & $\otimes$($\oplus$) & $\otimes$($\oplus$)& 
 $\otimes$($\oplus$) &
 $\otimes$($\oplus$) & $\otimes$($\oplus$) & $\otimes$($\oplus$) & $\otimes$($\oplus$)& 
 $\otimes$($\oplus$) & $\otimes$($\oplus$) & $\otimes$($\oplus$) 
\\
\hline 
\multirow{3}{*}{$Planck$} & $10^4$ $\sigma(\d n_s/ \d \ln k)$ & 59(60) & 53(55) & 60(60) & 57(58) & 59(60) & 53(57) & 60(60) & 56(57) & 60(60)  & 56(57) & 59(60) & 53(55)  \\

& $\sigma(f_{\rm NL})$& \bf{49(146)} & \bf{9.1(11)} & \bf{35(46)} & \bf{15(16)} & \bf{22(319)} & \bf{4.6(9.9)} & \bf{627(1443)} & \bf{10(11)} & \bf{16(277)} & \bf{2.1(2.8)}  & \bf{15(119)} & \bf{1.7(2.2)}  \\

& FoM & 3.5(1.1) & 21(18) & 4.8(3.6) & 12(11) & 7.7(0.5) & 41(19) & 0.3(0.1) & 18(17) & 11(1) & 86(65)  & 11(1) & 111(85)  \\

\hline 
\multirow{3}{*}{$Planck$+SO} &   $10^4$ $\sigma(\d n_s/ \d \ln k)$ & 31(31) & 30(30) & 31(31) & 31(31) &  31(31) & 30(31) & 31(31) & 31(31) & 31(31) & 31(31)  & 31(31) & 30(30)  \\

& $\sigma(f_{\rm NL})$& \bf{47(146)} & \bf{8.7(10)} & \bf{33(46)} & \bf{14(15)} &  \bf{22(319)} & \bf{4.5(9.3)} & \bf{618(1443)} & \bf{9.8(10)} & \bf{15(277)} & \bf{2.1(2.7)}  & \bf{14(120)} & \bf{1.7(2.1)}  \\

& FoM & 6.7(2.2) & 39(33) & 9.4(7.0) & 24(22) & 14.6(1.0) & 74(36) & 0.5(0.2) & 34(24) & 21(1.2) & 156(122)  & 22(2.7) & 201(159)  \\

\hline 
\multirow{3}{*}{LiteBIRD+S4} &  $10^4$ $\sigma(\d n_s/ \d \ln k)$ & 23(23) & 22(23) & 23(23) & 23(23) & 23(23) & 22(23) & 23(23) & 23(23) & 23(23)  & 23(23) & 23(23) & 22(22) \\

& $\sigma(f_{\rm NL})$ & \bf{42(146)} & \bf{8.2(9.9)} & \bf{31(46)} & \bf{13(14)} & \bf{19(319)} & \bf{4.1(9.0)} &  \bf{568(1443)} & \bf{9.4(9.8)} & \bf{12(207)} & \bf{2.0(2.7)}  & \bf{11(120)} & \bf{1.6(2.1)}  \\

& FoM & 10(3.0) & 55(46) & 14(9.5) & 34(32) & 23(1.4) & 113(50) & 0.8(0.3) & 47(46) & 34(1.6) & 224(166)  & 39(3.6) & 287(214)  \\

\hline 

\end{tabular}}
\label{tab:primordial}
\end{table}

\begin{table}
\scriptsize
\caption{68\% marginalized constraints on the 12 parameters of the extCDM cosmology obtained from the combination of each pair of CMB and galaxy surveys, in the single bin and tomographic configurations. We show the constraints from the full combination including the CMB-GC cross-correlation (CMB $\otimes$ GC) and, between parenthesis, the constraints from the uncorrelated combination of Fisher matrices (CMB $\oplus$ LSS). We also list the derived uncertainties on $\sigma_8$, and the FoMs of the 12 cosmological parameters and the bias nuisance parameters. For each survey combination, we mark in boldface the cosmological parameter which benefits most from including the CMB-LSS cross-correlation.}
\resizebox{\textwidth}{!}{
\begin{tabular}{|l|l|cccccccccccc|} 
\hline
\multicolumn{2}{|c}{extCDM} &  
\multicolumn{2}{|c}{Euclid-ph-like} &  \multicolumn{2}{c}{Euclid-sp-like} &
\multicolumn{2}{c}{LSST} &  \multicolumn{2}{c}{SPHEREx} &
\multicolumn{2}{c}{EMU} & \multicolumn{2}{c|}{SKA1}\\
\hline
\multirow{1}{*}{CMB} & \multirow{2}{*}{Parameter} & 1 bin & 10 bins & 1 bin & 9 bins & 1 bin & 10 bins & 1 bin & 10 bins & 1 bin & 5 bins & 1 bin & 5 bins \\

 \multirow{1}{*}{survey}&  &  $\otimes$($\oplus$) & $\otimes$($\oplus$) & $\otimes$($\oplus$) & $\otimes$($\oplus$) & 
 $\otimes$($\oplus$) & $\otimes$($\oplus$) & $\otimes$($\oplus$) & $\otimes$($\oplus$) &
 $\otimes$($\oplus$)& $\otimes$($\oplus$) & $\otimes$($\oplus$) & $\otimes$($\oplus$) 
\\
\hline 
\multirow{14}{*}{$Planck$} & $10^5$ $\sigma(\Omega_b h^2)$ & 25(25) & 19(20) & 25(25) & 22(22) & 24(25) & 19(21) & 25(25) & 19(22) & 24(25) & 21(23) & 24(25) & 19(21)  \\

& $10^4$ $\sigma(\Omega_c h^2)$
& 39(39) & 31(31) & 39(39) & 37(37) & 39(39) & 35(35) & 39(39) & 36(36) & 39(39) & 37(38) & 39(39) & 36(36) \\

& $\sigma(H_0)$ & 11(14) & 1.5(1.6) & 12(14) & \bf{9.1(10)} & 11(14) & 3.8(4.7) & 14(14) & 2.7(3.0) & 11(14) & \bf{3.9(6.4)} & 7.8(13) & \bf{3.6(6.6)} \\

& $10^4$ $\sigma(\tau)$ & 79(79) & 76(76) & 79(79) & 77(77) & 79(79) & 76(77) & 79(79) & 76(77) & 79(79) & 77(77) & 78(79) & 76(77)   \\

& $10^4$ $\sigma(n_s)$ & 109(110) & 80(86) & 109(110) & 97(97) & 109(110) & 83(91) & 109(110) & 87(96) & 108(110) & 91(103) & 108(110) & 85(95) \\

& $10^3$ $\sigma(\ln (10^{10} A_s))$ & 18(18) & 17(17) & 18(18) & 17(17) & 18(18) & 17(17) & 18(18) & 17(17) & 18(18) & 17(17) & 18(18) & 17(17)  \\

& $10^2$ $\sigma(w_0)$& 77(94) & 11(11) & 88(95) & 72(79) & 80(94) & 33(40) & 88(94) & 19(20) & 80(93) & 33(48) & 68(93) & 29(50)  \\

& $10^2$ $\sigma(w_a)$& 202(214) & 37(41) & 211(214) & 164(172) & 198(211) & 86(106) & 193(211) & 43(50) & 190(208) & 62(92) & 174(210) & 71(113) \\

& $\sigma(\Sigma m_\nu)$/meV & 229(265) & 122(139) & 218(263) & 181(181) & 192(223) & 128(164) & 221(224) & \bf{87(178)} & 179(223) & 129(187) & 170(222) & 109(143)  \\

& $10^2$ $\sigma(N_{\rm eff})$ & 27(27) & 21(22) & 27(27) & 25(25) & 27(27) & 23(24) & 27(27) & 24(25) & 27(27) & 25(26) & 27(27) & 23(24)   \\

& $10^4$ $\sigma(\d n_s/ \d \ln k)$ & 75(76) & 64(66) & 74(76) & 70(70) & 74(75) & 65(69) & 75(75) & 67(69) & 75(75) & 68(71) & 74(75) & 66(69) \\

& $\sigma(f_{\rm NL})$& \bf{49(146)} & \bf{9.2(11)} & \bf{35(46)} & 16(17) & \bf{22(320)} & \bf{4.7(10)} & \bf{645(1443)} & 10(10) & \bf{16(282)} & 2.2(2.9) & \bf{16(139)} & 1.7(2.3) \\

& $10^3$ $\sigma(\sigma_8)$& 96(121) & 16(17) & 103(121) & 80(88) & 96(117) & 35(44) & 113(117) & 25(27) & 95(116) & 36(57) & 69(115) & 33(59)  \\

& FoM$_{\rm cosmo}$ / $10^2$ & 8(6) & 29(26) & 9(8) & 14(14) & 10(6) & 24(18) & 5(4) & 26(22) & 11(6) & 25(18) & 12(7) & 29(21)\\

& FoM$_{\rm bias}$ / $10^2$ & 0.11(0.02) & 439(405) & 0.011(0.004) & 22(22) & 0.3(0.02) & 43(32) & 0.017(0.001) & 35(30) & 1.9(0.7) & 55(35) & 0.4(0.2) & 60(37) \\

\hline 
\multirow{14}{*}{$Planck$+SO} & $10^5$ $\sigma(\Omega_b h^2)$ & 7.9(7.9) & 7.5(7.7) & 7.9(7.9) & 7.9(7.9) & 7.9(7.9) & 7.5(7.8) & 7.9(7.9) & 7.6(7.8) & 7.9(7.9) & 7.7(7.9) & 7.9(7.9) & 7.4(7.8) \\

& $10^4$ $\sigma(\Omega_c h^2)$
& 15(15) & 14(14) & 15(15) & 15(15) & 15(15) & 14(14) & 15(15) & 14(14) & 15(15) & 14(14) & 15(15) & 14(14)  \\

& $\sigma(H_0)$ & 3.7(3.7) & 1.1(1.1) & 3.6(3.7) & 3.3(3.3) & 3.6(3.7) & 2.0(2.7) & 3.7(3.7) & 1.9(2.2) & 3.7(3.7) & \bf{2.3(3.2)} & 3.1(3.7) & \bf{1.9(3.2)}  \\

& $10^4$ $\sigma(\tau)$ &75(75) & 71(72) & 75(75) & 73(74) & 75(75) & 71(73) & 75(75) & 68(72) & 75(75) & 73(75) & 75(75) & 72(74) \\

& $10^4$ $\sigma(n_s)$ & 41(41) & 39(39) & 41(41) & 40(40) & 41(41) & 38(40) & 41(41) & 39(40) & 41(41) & 40(40) & 41(41) & 38(40)  \\

& $10^3$ $\sigma(\ln (10^{10} A_s))$ & 15(15) & 14(14) & 15(15) & 14(14) & 15(15) & 14(14) & 15(15) & 13(14) & 15(15) & 15(15) & 15(15) & 14(15)  \\

& $10^2$ $\sigma(w_0)$& 26(27) & 9.6(10) & 27(27) & 26(26) & 26(27) & 18(22) & 27(27) & 14(15) & 27(27) & 20(24) & 26(27) & 16(24)\\

& $10^2$ $\sigma(w_a)$& 75(80) & 32(36) & 80(80) & 67(68) & 75(80) & 51(62) & 79(80) & 34(38) & 78(80) & 58(69) & 78(80) & 47(67)  \\

& $\sigma(\Sigma m_\nu)$/meV & 90(92) & \bf{69(80)} & 92(92) & 88(89) & 89(92) & 71(87) & 92(92) & \bf{44(86)} & 92(92) & 78(89) & 90(92) & 68(85) \\

& $10^2$ $\sigma(N_{\rm eff})$ & 8.9(8.9) & 8.5(8.6) & 8.9(8.9) & 8.8(8.8) & 8.9(8.9) & 8.5(8.8) & 8.9(8.9) & 8.8(8.8) & 8.9(8.9) & 8.8(8.8) & 8.9(8.9) & 8.5(8.7)  \\

& $10^4$ $\sigma(\d n_s/ \d \ln k)$ & 44(44) & 42(42) & 44(44) & 43(43) & 44(44) & 42(43) & 44(44) & 43(43) & 44(44) & 43(43) & 44(44) & 42(43)   \\

& $\sigma(f_{\rm NL})$& \bf{47(145)} & 8.8(10) & \bf{33(45)} & \bf{14(15)} & \bf{22(319)} & \bf{4.5(9.5)} & \bf{622(1443)} & 10(10) & \bf{15(278)} & 2.1(2.8) & \bf{14(121)} & 1.7(2.2)   \\

& $10^3$ $\sigma(\sigma_8)$& 31(31) & 9(10) & 31(31) & 28(29) & 31(31) & 19(24) & 31(31) & 17(18) & 31(31) & 20(28) & 27(31) & 17(28) \\

& FoM$_{\rm cosmo}$ / $10^2$ & 40(33) & 87(79) & 42(41) & 55(55) & 46(29) & 84(64) & 26(23) & 85(74) & 48(30) & 83(70) & 51(35) & 98(76)  \\

& FoM$_{\rm bias}$ / $10^2$ & 0.50(0.02) & 589(504) & 0.034(0.004) & 29(28) & 1.09(0.02) & 72(43) & 0.028(0.001) & 40(35) & 7.0(1.2) & 92(65) & 1.6(0.4) & 108(68)  \\

\hline

\multirow{14}{*}{LiteBIRD+S4} & $10^5$ $\sigma(\Omega_b h^2)$ & 4.6(4.6) & 4.4(4.5) & 4.6(4.6) & 4.6(4.6) & 4.6(4.6) & 4.4(4.6) & 4.6(4.6) & 4.5(4.6) & 4.6(4.6) & 4.4(4.6) & 4.6(4.6) & 4.4(4.6)  \\

& $10^4$ $\sigma(\Omega_c h^2)$
& 10(10) & 9.9(9.9) & 10(10) & 10(10) & 10(10) & 10(10) & 10(10) & 9.8(10) & 10(10) & 10(10) & 10(10) & 10(10) \\

& $\sigma(H_0)$ & 2.0(2.0) & 0.9(1.0) & 2.0(2.0) & 1.8(1.8) & 2.0(2.0) & 1.3(1.7) & 2.0(2.0) & 1.4(1.5) & 2.0(2.0) & 1.5(1.9) & 1.8(2.0) & 1.3(1.8) \\

& $10^4$ $\sigma(\tau)$ & 18(18) & 18(18) & 18(18) & 18(18) & 18(18) & 18(18) & 18(18) & 18(18) & 18(18) & 18(18) & 18(18) & 18(18)   \\

& $10^4$ $\sigma(n_s)$ & 29(29) & 27(28) & 29(29) & 28(28) & 29(29) & 26(28) & 29(29) & 28(28) & 29(29) & 28(28) & 29(29) & 27(28)  \\

& $10^3$ $\sigma(\ln (10^{10} A_s))$ & 4.1(4.2) & 4.0(4.1) & 4.2(4.2) & 4.1(4.1) & 4.1(4.2) & 4.0(4.1) & 4.2(4.2) & 3.8(4.1) & 4.2(4.2) & 4.1(4.2) & 4.1(4.2) & 4.0(4.1) \\

& $10^2$ $\sigma(w_0)$& 15(15) & 8.2(9.2) & 15(15) & 14(14) & 15(15) & 12(14) & 15(15) & 11(11) & 15(15) & 13(14) & 15(15) & 11(14) \\

& $10^2$ $\sigma(w_a)$& 48(52) & 27(32) & 53(53) & 45(46) & 46(53) & 34(43) & 52(52) & 27(30) & 50(53) & 41(49) & 50(53) & 33(47) \\

& $\sigma(\Sigma m_\nu)$/meV & 54(56) & \bf{38(51)} & 57(57) & 54(56) & 53(57) & 37(54) & 58(56) & \bf{28(51)} & 56(57) & 46(56) & 55(57) & \bf{36(55)} \\

& $10^2$ $\sigma(N_{\rm eff})$ & 5.9(5.9) & 5.8(5.9) & 5.9(5.9) & 5.9(5.9) & 5.9(5.9) & 5.7(5.9) & 5.9(5.9) & 5.9(5.9) & 5.9(5.9) & 5.9(5.9) & 5.9(5.9) & 5.8(5.9) \\

& $10^4$ $\sigma(\d n_s/ \d \ln k)$ & 34(34) & 33(33) & 34(34) & 34(34) & 34(34) & 32(34) & 34(34) & 33(33) & 34(34) & 34(34) & 34(34) & 33(34) \\

& $\sigma(f_{\rm NL})$& \bf{42(145)} & 8.3(10) & \bf{31(45)} & \bf{14(15)} & \bf{19(319)} & \bf{4.1(9.2)} & \bf{569(1443)} & 9.7(10) & \bf{13(277)} & \bf{2.0(2.7)} & \bf{11(119)} & 1.6(2.1) \\

& $10^3$ $\sigma(\sigma_8)$& 17(17) & 7(8) & 17(17) & 15(16) & 17(17) & 12(14) & 17(17) & 11(13) & 17(17) & 13(16) & 15(17) & 11(15)  \\

& FoM$_{\rm cosmo}$ / $10^2$ & 109(88) & 203(177) & 112(106) & 139(138) & 128(77) & 216(155) & 69(60) & 199(172) & 132(79) & 207(177) & 139(90) & 245(189)  \\

& FoM$_{\rm bias}$ / $10^2$ & 0.93(0.02) & 759(563) & 0.059(0.005) & 33(32) & 2.03(0.02) & 108(49) & 0.038(0.001) & 45(39) & 12(1.3) & 132(83) & 4.1(0.4) & 172(92)   \\

\hline 

\end{tabular}}
\label{tab:extCDM}
\end{table}

 \clearpage
 
 \section{Scale-dependent bias induced by massive neutrinos}
 
 \label{sec:neutrino}

In this appendix we quantify the impact of a scale dependence of the galaxy bias induced by a neutrino mass, which is an additional effect which has drawn a lot of attention recently \cite{LoVerde:2014pxa,Raccanelli:2017kht,Vagnozzi:2018pwo,Chiang:2018laa}, but has not been considered as a baseline in the body of the paper. 
 
In order to estimate its impact on uncertainties of cosmological parameters, we consider the scale dependence of the galaxy bias as in \cite{Tanidis:2020byi}, i.e. as a smooth transition around $k_{\rm nr}$, which is the free streaming scale for the non-relativistic neutrinos. This modelling of the scale-dependent bias is given by \cite{Tanidis:2020byi}:
\begin{equation}
    b(k,z) = b_{k \ll k_{\rm nr}}(z) + \frac{b_{k \ll k_{\rm nr}}(z)-b_{k \gg k_{\rm nr}}(z)}{2}
    \left\{ \tanh \left[ \ln \left(\frac{k}{k_{\rm nr}} \right)^\gamma \right] +1 \right\}
\end{equation}
 where $k_{\rm nr} \approx 0.018 (\frac{m_\nu}{1 {\rm eV}})^{1/2} \sqrt{\Omega_{\rm m}} h$  Mpc$^{-1}$ and $\gamma = 5$. The values of $b_{k \ll k_{\rm nr}}(z)$ and $b_{k \gg k_{\rm nr}}(z)$ correspond to the two asymptotic regimes of the bias: at $k$ much smaller than $k_{\rm nr}$ we recover the standard $\Lambda$CDM galaxy bias, hence $b_{k \ll k_{\rm nr}}(z) \simeq b_G(z)$; while at $k$ larger than $k_{\rm nr}$ the bias corresponds to $b_{k \gg k_{\rm nr}}(z) \simeq b_G(z)(1-f_{\nu})$, where $f_{\nu} = \Omega_\nu / \Omega_{\rm m}$. 
 
 In order to take into account surveys of different redshift coverage and depth, we implement the scale-dependent bias and recompute the constraints on the $\Lambda$CDM+$\{\Sigma m_\nu,N_{\rm eff} \}$ cosmology for the Euclid-ph-like, SPHEREx and SKA1 surveys using their tomographic configurations. We list in Tab.~\ref{tab:bias} the constraints on the neutrino mass compared to the case neglecting the scale-dependent bias for these three surveys, alone and in their uncorrelated and full combinations with the $Planck$-like CMB survey.

 \begin{table}[h]
     \centering
     \begin{tabular}{|c|c|c|c|} \hline 
     & Euclid-ph-like & SPHEREx & SKA1 \\ \hline
     $C_\ell^{GG}$  & 297 (347) & 510 (512) & 407 (575) \\ \hline
     $Planck \oplus C_\ell^{GG}$  & 67 (68) &  61 (61) & 92 (107)  \\ \hline
     $Planck \otimes C_\ell^{GG}$  & 72 (73)  & 55 (55) & 77 (85)  \\ \hline

     \end{tabular}
     \caption{Uncertainty on the neutrino mass $\sigma(m_\nu)$ in meV for each galaxy survey and their uncorrelated and full combinations with the $Planck$ CMB survey, obtained after the implementation of the scale-dependent bias. The numbers between parenthesis correspond to the uncertainty calculated neglecting the scale-dependent bias.}
     \label{tab:bias}
 \end{table}

We find tighter constraints in the neutrino mass from galaxy surveys after having introduced the scale-dependent bias. For galaxy clustering only, the error is smaller by less than 1\% for SPHEREx, but this difference grows to $\sim 15\%$ for Euclid-ph-like and $\sim 30\%$ for SKA1. This suggests that the effect of the scale-dependent bias is more important for deeper surveys in redshift. When combining (including its cross-correlation) with the $Planck$ information, these differences are $\sim 14\%$ ($\sim 10\%$) for SKA1 and $\lesssim 2\%$ for Euclid-ph-like. For the combination with more powerful CMB surveys such as SO and S4, these differences should be even smaller. We therefore conclude that our results in Sect.~\ref{sec:results} based on linear scales are robust to the inclusion of the scale-dependence in the galaxy bias due to neutrino mass, an additional effect which can actually slightly decrease the expected uncertainty on the neutrino mass.


\end{document}